\documentclass[preprint,12pt]{aastex}

\usepackage[usenames,dvipsnames]{color}
\usepackage{amsmath,rotating,listings,xfrac,url,float,natbib}

\newcommand{\mm}{\ensuremath{\mathcal{M}}}
\newcommand{\msun}{\ensuremath{\mathcal{M}_{\odot}}}
\newcommand{\rr}{\ensuremath{\mathcal{R}}}
\newcommand{\rsun}{\ensuremath{\mathcal{R}_{\odot}}}

\newcommand{\kms}{\ensuremath{{\rm ~km~s}^{-1}}}
\makeatletter
\newcommand{\Rmnum}[1]{\expandafter\@slowromancap\romannumeral #1@}

\newcommand{\kep}{{\it Kepler }}
\begin{document}

\title{WOCS 40007: A Detached Eclipsing Binary near the Turnoff of the Open
  Cluster NGC 6819\footnote{This is paper 56 of the WIYN Open Cluster Study
    (WOCS).}}  \author{Mark W. Jeffries, Jr.\altaffilmark{2}; Eric
  L. Sandquist\altaffilmark{2}; Robert D. Mathieu\altaffilmark{3,11}; Aaron
  M. Geller\altaffilmark{3,4,11}; Jerome A. Orosz\altaffilmark{2}; Katelyn
  E. Milliman\altaffilmark{3,11}; Lauren N. Brewer\altaffilmark{2}; Imants
  Platais\altaffilmark{5}; Karsten Brogaard\altaffilmark{6,7}; Frank
  Grundahl\altaffilmark{7}; Soeren Frandsen\altaffilmark{7}; Aaron
  Dotter\altaffilmark{8,9}; Dennis Stello\altaffilmark{10}}

\altaffiltext{2}{San Diego State University, Department of Astronomy,
  San Diego, CA, 92182; {\tt jeffries@sciences.sdsu.edu}; {\tt
    erics@mintaka.sdsu.edu}; {\tt orosz@sciences.sdsu.edu}; {\tt
    lbrewer@rohan.sdsu.edu}} 
\altaffiltext{3}{University of Wisconsin-Madison, Department of
  Astronomy, Madison, WI, 53706; {\tt mathieu@astro.wisc.edu}; {\tt
    milliman@astro.wisc.edu}}
\altaffiltext{4}{Center for Interdisciplinary Exploration and Research 
   in Astrophysics (CIERA) \& Dept. of Physics and Astronomy, 
   Northwestern University, 2145 Sheridan Road, Evanston, IL 60208
   {\tt a-geller@northwestern.edu}}    
\altaffiltext{5}{Department of Physics and Astronomy, The Johns Hopkins University, Baltimore, MD 21218, USA ; {\tt imants@pha.jhu.edu}}
\altaffiltext{6}{Department of Physics \& Astronomy, University of Victoria,
  P.O. Box 3055, Victoria, BC V8W 3P6, Canada}
\altaffiltext{7}{Department of Physics and Astronomy, Aarhus University, Ny
  Munkegade 120, 8000 Aarhus C, Denmark; {\tt kfb@phys.au.dk, fgj@phys.au.dk,
    srf@phys.au.dk}}
\altaffiltext{8}{Research School of Astronomy and Astrophysics, The Australian
  National University, Canberra, ACT, Australia; {\tt aaron.dotter@gmail.com}}
\altaffiltext{9}{Space Telescope Science Institute, 3700 San Martin Drive,
  Baltimore, MD, 21218}
\altaffiltext{10}{Sydney Institute for Astronomy (SIfA), School of Physics,
  University of Sydney, NSW, 2006, Australia; {\tt stello@physics.usyd.edu.au}}
\altaffiltext{11}{Visiting Astronomer, Kitt Peak National Observatory, 
  National Optical Astronomy Observatories, which is operated by the 
  Association of Universities for Research in Astronomy, Inc. (AURA) under 
  cooperative agreement with the National Science Foundation.}
\begin{abstract}

We analyze extensive $BVR_cI_c$ time-series photometry and
radial-velocity measurements for WOCS 40007 (Auner 259; KIC 5113053), a
double-lined detached eclipsing binary and a member of the open cluster NGC
6819.  Utilizing photometric observations from the 1-meter telescope at Mount
Laguna Observatory and spectra from the WIYN 3.5-meter telescope, we measure
precise and accurate masses ($\sim1.6$\% uncertainty) and radii ($\sim0.5$\%)
for the binary components.  In addition, we discover a third star orbiting the
binary with a period greater than 3000 days using radial velocities and \kep
eclipse timings.

Because the stars in the eclipsing binary are near the cluster turnoff, they
are evolving rapidly in size and are sensitive to age.  With a metallicity of
[Fe/H]$=+0.09\pm0.03$,
we find the age of NGC 6819 to be about 2.4 Gyr from CMD isochrone fitting and
$3.1\pm0.4$ Gyr by analyzing the mass-radius ($M-R$) data for this binary. The
$M-R$ age is above previous determinations for this cluster, but consistent
within $1\sigma$ uncertainties. When the $M-R$ data for the primary star of the
additional cluster binary WOCS 23009 is included, the weighted age estimate 
drops to $2.5\pm0.2$ Gyr, with a systematic uncertainty of at least 0.2 Gyr.
The age difference between our CMD and $M-R$ findings may be the result of 
systematic error in the metallicity or helium abundance used in models, or 
due to slight radius inflation of one or both stars in the WOCS 40007 binary.
\end{abstract}

\keywords{open clusters and associations: individual (NGC 
6819) - stars: evolution - stars: binaries: spectroscopic 
- stars: binaries: eclipsing - techniques: spectroscopy - 
techniques: photometry}

\section{Introduction}\label{intro}
Detached eclipsing binaries (DEBs) provide a means of deriving precise
stellar ages while avoiding most of the major systematic errors in
isochrone fits that are introduced by uncertainties in model
assumptions, reddening, distance, and color-temperature
transformations \citep{andersen,torres}.  The masses and radii of the
stellar components in such a binary make a {\it direct} comparison
possible with stellar evolution theory if chemical composition is
known independently. The characteristics of the stars (especially
radius) will also have sensitivity to age if one or both of the stars
has evolved significantly away from the main sequence.  If an
eclipsing binary containing such an evolved star can be found in a
star cluster, it can place tight constraints on the age of {\it all}
of the cluster stars.

With the start of NASA's \kep mission, star clusters within its field of view
are becoming important test beds for stellar evolution theories. \kep has
discovered previously unknown eclipsing binary systems 
both in the field \citep{kebs1,kebs2} and in open clusters
\citep[e.g.,][]{sandquist13}. As the population of binary members with known
masses and radii increases, it will be possible to statistically improve the
binary star constraint on cluster ages derived from the mass-radius ($M-R$)
plane. Additionally, a well-populated $M-R$ plane
observationally constrains difficult-to-measure
cluster parameters, such as He abundance, when analyzed together with  a cluster
CMD \citep{grundahl,brogaard,brogaard12}.

NGC 6819 is a rich open cluster in the \kep field that has been 
well studied.  \citet{hole} completed the first comprehensive kinematic
membership study as a part of the WIYN Open Cluster Study
\citep[WOCS;][]{mathieu}, including stars from the upper main sequence (MS)
to the giant branch.  Several previous studies of the color-magnitude
diagram \citep[CMD;][]{burkhead,lindoff,auner,rosvick,kalirai,kalirai04} have
produced age estimates between about 2.0 and 2.5 Gyr.  More recently,
\citet{basu} used asteroseismic data from \kep to constrain
characteristics of red giant branch stars and to place a constraint on the
cluster distance and age. They find $(m-M)_0 = 11.85\pm0.05$ and an age of
$2.1-2.5$ Gyr.

Our group has undertaken a program to analyze eclipsing binary stars in this
cluster, and to improve the determination of its age using the {\it ensemble}
of eclipsing systems. In our first paper \citep{sandquist13}, we analyzed the
long-period binary WOCS 23009 (Auner
851), which contains the most evolved star known to eclipse in this
cluster. Here we focus on the binary system WOCS 40007 (Auner 259;
$\alpha_{2000}=19^{\rm h}41^{\rm m}33\farcs94$,
$\delta_{2000}=+40\degr13\arcmin00\farcs5$), which was discovered and
classified as a DEB by \citet{talamantes}. WOCS 40007 contains stars that are
less evolved than the primary star in WOCS 23009, and so they are less age
sensitive. However, both stars can contribute to the determination of the
cluster age, and they also allow us to measure the characteristics of stars of
other masses in the cluster. This study uses
  extensive ground-based photometric and radial-velocity (RV) observations to
  characterize the binary, and \kep photometry to monitor eclipse timing and
  examine out-of-eclipse variations.

Section \ref{metred} contains a discussion of the metallicity, reddening, and
distance of NGC 6819.  A discussion of the photometric and spectroscopic
observations of the binary star, as well as the reduction and modeling of the
data, is described in \S \ref{obs}.  In \S \ref{obsresults} we present our
photometric and spectroscopic results for the components of WOCS 40007.  
Finally, we determine the age for NGC 6819 and discuss our results in 
\S \ref{iso}.

\section{Physical Parameters of NGC 6819}\label{metred}

Before describing the observations, we review
important external information that goes into later analysis: metallicity,
reddening, and distance.

\subsection{Metallicity}\label{met}

Uncertainty in the metallicity of cluster stars potentially produces one of
the biggest systematic uncertainties in age via the theoretical prediction of
radii.  \citet{friel} first determined the metallicity of NGC 6819 with a
study of moderate-resolution spectra of giant stars and found [Fe/H]$=-0.1$.
Later work with the same spectroscopic data led \citet{friel1993} to derive
[Fe/H]$=+0.05\pm0.11$.  However, both of these results are dependent on the
assumed reddening of the system and on data that the authors acknowledged to
be of high uncertainty.  \citet{twarog} found [Fe/H]$=+0.07\pm0.05$ using the
data of \citet{friel1993}, but with an independent estimate of the 
cluster reddening.  More recently, \citet{bragaglia} used high-dispersion 
spectra of three red clump stars to derive the cluster metallicity and 
element-to-element abundances for NGC 6819.  They found that their targets 
possess abundance ratios quite close to solar values except for a small 
excess of silicon and large overabundance of sodium.  \citet{bragaglia} find 
the cluster to be slightly more metal-rich than the Sun with 
[Fe/H]$=+0.09\pm0.03$.  To date, the \citet{bragaglia} value is the only
published metallicity using high resolution spectra,
and so we use this as our preferred value in 
the subsequent analysis. Although NGC 6819 is part of a larger homogeneous
study of open cluster abundances \citep{bocce}, there has not yet been a 
systematic comparison of their abundances with previous measurements in
well-studied clusters. As a result, we have to allow for the possibility
of a systematic error in [Fe/H] that could be larger than the quoted 
uncertainty in the mean.

\subsection{Reddening and Distance}\label{reddist}
Previously quoted reddening values in the literature range from $E(B-V)=0.12$
\citep{burkhead}, $0.30$ \citep{lindoff}, $0.28$ \citep{auner}, $0.15$
\citep{canterna}, and $0.142\pm0.044$ \citep{bragaglia}.  
In the most recent study, \citet{bragaglia} derived the cluster
reddening by determining the temperatures of 3 bright red clump stars
from line excitation (a reddening-free parameter) and then comparing
the observed color of each star with values predicted by theoretical
models. It should be noted that one of the stars observed by
\citeauthor{bragaglia} is a binary (Auner 979; \citealt{hole}), and so
its contribution is suspect. In order to make the most reliable
comparisons between results from the color-magnitude diagram and from
the components of WOCS 40007 (which are independent of distance and
reddening), we redetermine the reddening here.

To compute the reddening, we used a method based on \citet{grocholski},
comparing the median red clump photometry of NGC 6819 with that of M67, a
cluster with low and well-determined reddening
\citep[$E(B-V)=0.041\pm0.004$;][]{taylor} and distance modulus
\citep[$(m-M)_V=9.72\pm0.05,~(m-M)_0=9.60\pm0.03$;][]{sandquist04}. M67 and
NGC 6819 also have similar ages and metallicities that fall in a regime where
variations in those quantities have modest effects on the brightness of the
red clump. The metallicity difference $\Delta$[Fe/H]$=0.09$ is theoretically
expected to affect the clump magnitude at only the 0.01 level in 2MASS
infrared filters, but has a larger effect towards bluer wavelengths 
\citep{girardi}, reaching almost 0.1 mag in $B$.
The $\sim1.5$ Gyr age difference between the two clusters implies that the M67
red clump should be fainter in all filters --- approximately 0.07 mag in
2MASS filters and increasing to about 0.14 mag in $B$
\citep{girardi,grocholski}.
As a result, for these two clusters the age and
metallicity effects are theoretically expected to partially compensate for
each other, so that there is a nearly constant magnitude offset in all
filters. Based on these considerations, we can use the magnitude of the red
clump stars in different filter bands to seek a simultaneous solution for the
difference in the cluster distance moduli $\Delta (m-M)_0$ and difference in
optical depth to the clusters $\Delta \tau_1$ if an extinction law is assumed.
The difference in distance moduli is mostly set by magnitude differences in
the infrared where extinction effects are small, while the difference in
optical depth is determined by how the magnitude differences change from
filter to filter.

\begin{figure}
\begin{center}
\includegraphics[width=\textwidth]{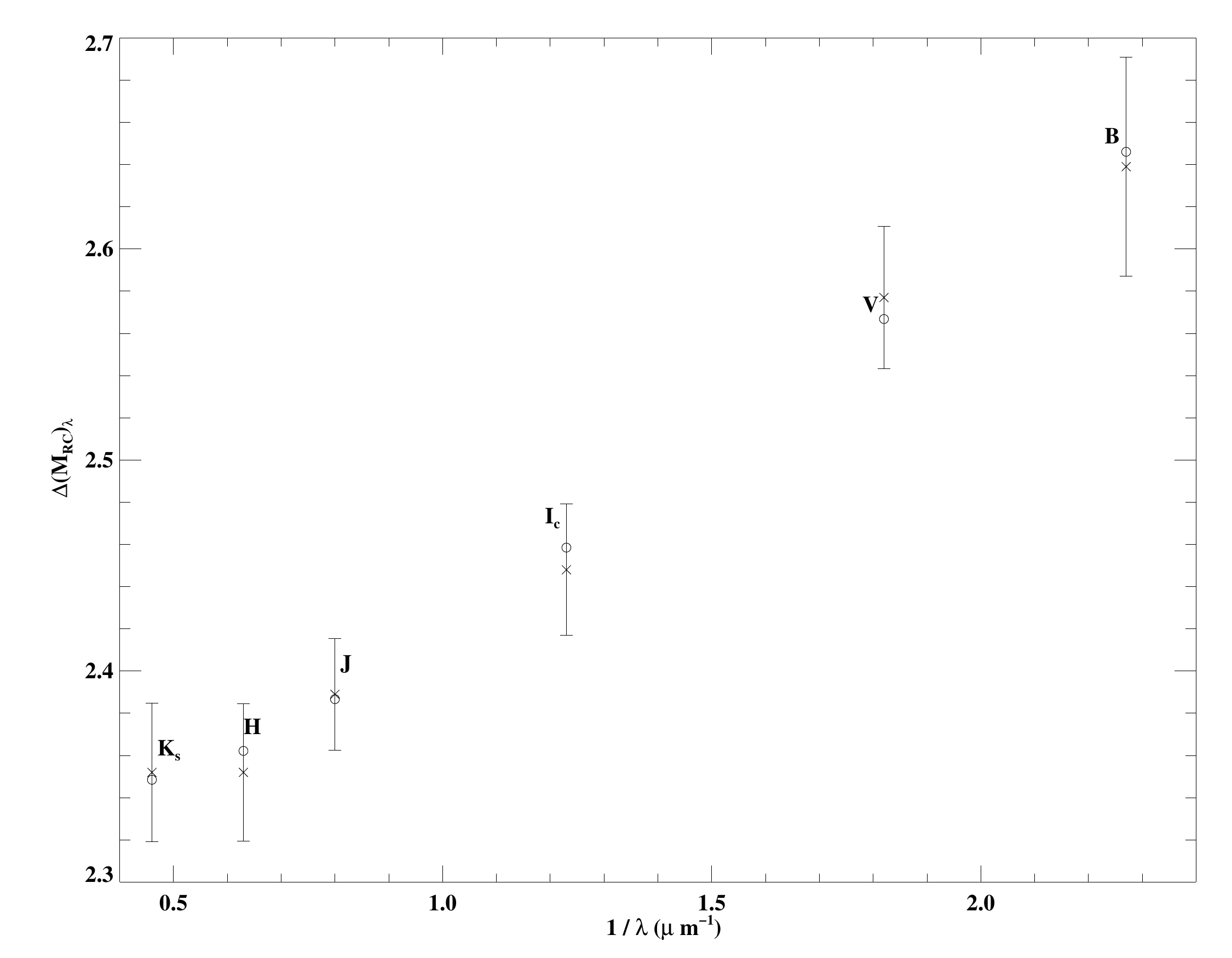}
\caption{ The difference in median clump magnitude in $B,~V,~I_c,~J,~H,$ and
  $K_s$, as a function of wave number ($1/\lambda$) between NGC 6819 and M 67.
  Crosses show measured values, while open circles indicate the best fit
  using the extinction algorithm of \citet{mccall}.}
\label{redcomp}
\end{center}
\end{figure}

We used $BVI_c$ (\citealt{talamantes} for NGC 6819, and \citealt{sandquist04}
for M67) and $JHK_s$ photometry \citep[2MASS]{2mass} for clump stars. In NGC
6819, we restrict the sample to single-star members as identified in
\citet{hole} and also eliminated likely evolved clump stars identified from
asteroseismology \citep{corsaro}.  We also employ the York Extinction
Solver\footnote{\tt{http://www2.cadc-ccda.hia-iha.nrc-cnrc.gc.ca/community/YorkExtinctionSolver/}}
\citep{mccall} to calculate the reddening and extinctions in different
filters, assuming that the clump stars are approximately K0 \Rmnum{3}
giants. We are able to get a consistent solution for all filter bands with
$\Delta (m-M)_0=2.40^{+0.03}_{-0.02}$ and $\Delta \tau_1 = 0.088\pm0.032$.
The uncertainties depend mostly on the uncertainties in the {\it relative}
abundance of M67 and NGC 6819, which we estimate to be about 0.05 dex. 
This estimate is based on the work of \citet{friel1993}, which is the only 
spectroscopic study to measure the metallicity of NGC 6819 and M67
in a consistent way that would minimize systematic errors.
This study found a difference of 0.14 dex, with NGC 6819 being the more 
metal-rich cluster.  After accounting for M67's distance and reddening, we 
get $(m-M)_0=12.00\pm0.05$, $(m-M)_V=12.37\pm0.10$ and $E(B-V)=0.12\pm0.03$ 
for dwarf stars near the turnoff of NGC 6819.  Figure \ref{redcomp} 
illustrates the consistency in median red clump magnitudes between the two 
clusters in $BVI_cJHK_s$ as a function of wave number ($1/\lambda$) when 
incorporating these shifts.

\begin{figure}
\begin{center}
\includegraphics[width=\textwidth]{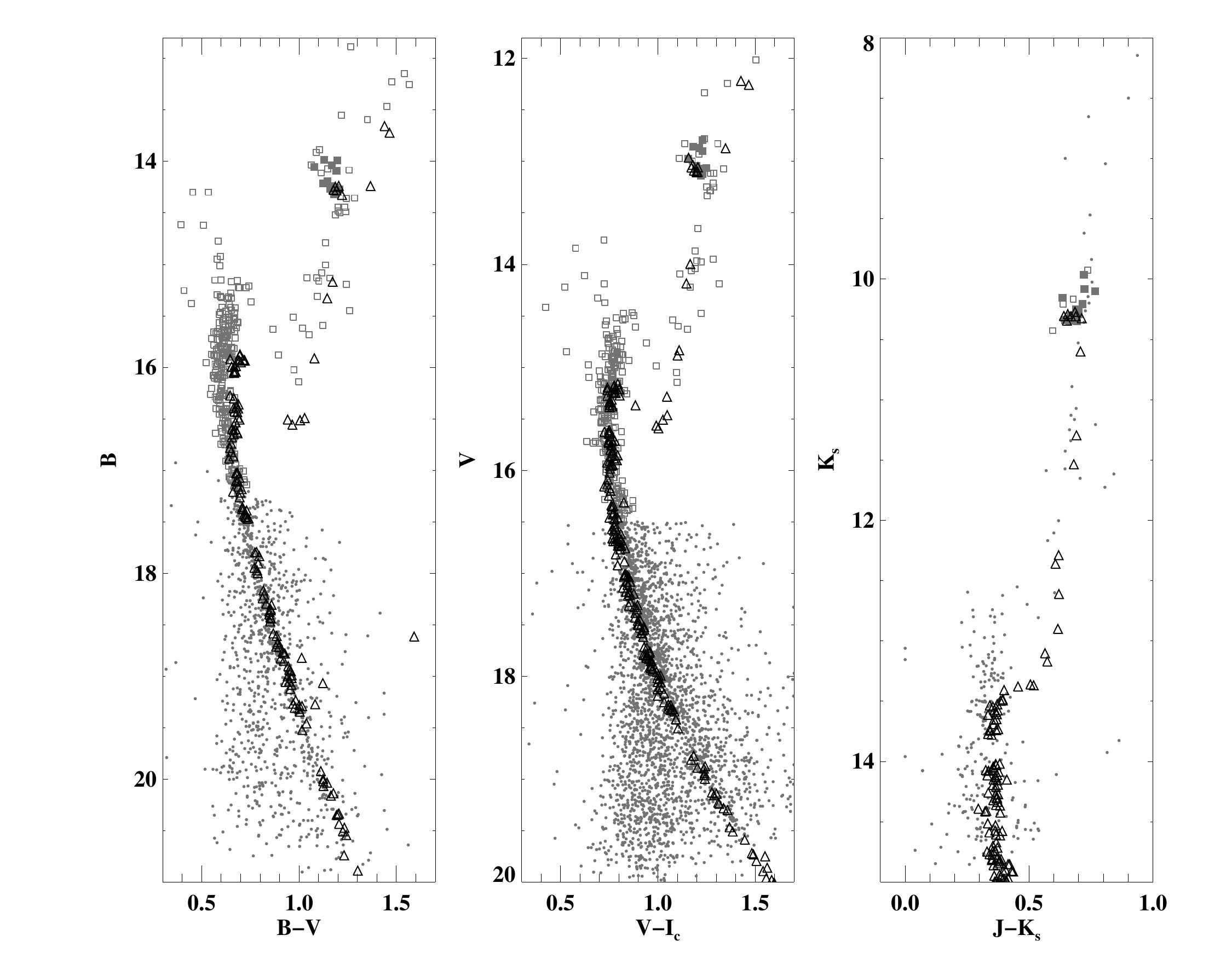}
\caption{
Comparison of CMDs of M67 and NGC 6819. Single-star members
of M67 \citep{sandquist04} are plotted as open triangles.
M67 stars have been shifted in $B$, $V$ and $K_s$ magnitude 
by $2.646$, $2.567$, and $2.349$, respectively.  Color shifts
of $\Delta(B-V)=0.079$, $\Delta(V-I_c)=0.108$, and 
$\Delta(J-K_s)=0.038$ mag are also applied.
Open grey squares represent single-star members of NGC 
6819  and the filled grey squares are the known 
red clump population \citep{hole}.  Grey dots with $V>16.5$ 
are from the full CMD of \citet{talamantes}.
}
\label{redcheck}
\end{center}
\end{figure}

The reddening is at the low end of the range of previous determinations
\citep{bragaglia,canterna,burkhead}.  Our value for the apparent distance
modulus agrees with the measurements by
\citet[][$(m-M)_V\approx12.35$]{rosvick} and
\citet[][$(m-M)_V=12.30\pm0.12$]{kalirai}, although larger than the recent
asteroseismic value determined by \citet[][$(m-M)_0=11.85\pm0.05$]{basu} using
a reddening $E(B-V)=0.15$. Figure \ref{redcheck} shows the consistency between
the MS and clumps in the resulting corrected CMDs of NGC 6819 and M67.  We
also note that the turnoff of M67 is fainter and redward of the NGC 6819
turnoff, indicating that NGC 6819 is younger, as expected. We will discuss
independent distance moduli derived from the eclipsing binary stars in \S
\ref{secdm}.

\section{Observations and Data Reduction}\label{obs}

\subsection{Photometry}\label{phot}

We obtained CCD images in $B,~V,~R_c$, and $I_c$ of NGC 6819 using the 1-m
telescope at Mount Laguna Observatory (hereafter, MLO).  The CCD 2005 camera
covers $13.7^\prime\times 13.7^\prime$ on the sky, resulting in a scale of
0\farcs41 per pixel.  Our photometric dataset consists of 46 nights between
March 2001 and September 2011, with the majority from 2008, 2010, and
2011. The dataset includes images presented by \citet{talamantes} that were
taken with the same telescope and camera combination. A list of additional
nights of photometry is in Table \ref{tableobs}.

Our data reduction pipeline uses {\it IRAF}\footnote{{\it
    IRAF} is distributed by the National Optical Astronomy Observatory, which
  is operated by the Association of Universities for Research in Astronomy,
  Inc., under cooperative agreement with the National Science Foundation.}
routines for overscan and bias subtraction, and flat fielding.  Our science
images are flat-field corrected using images acquired during the twilight of
each night, although for a few nights we use a flat-field image composed from
dome and twilight flats, as noted by a superscript ``b'' in Table
\ref{tableobs}.

We use a slightly modified version of the image subtraction package
ISIS \citep{alard} to make photometric measurements, as described in
\citet{sandquist12} and \citet{talamantes}.  In short, we improved
the handling of image alignment, fixed an error in the uncertainty
determination, and modified how ISIS acquires the reference flux in a
crowded field.  Using ISIS, we interpolate our $B, ~V,~R_c$, and $I_c$
data sets to a common coordinate system.  Reference images are made
for each filter from about 20 of the best seeing images, and these
were subtracted from individual science images after being convolved
to match the seeing.

The output of ISIS is a measurement of the difference flux from the subtracted
images.  The ISIS aperture photometry routine uses a circular aperture of {\tt
  radphot} pixels, weighted by the point-spread function and normalized to a
larger aperture with a radius of {\tt rad\_aper} pixels.  For our MLO
datasets, we set {\tt radphot} $=4$ and {\tt rad\_aper} $=7$.  The magnitude
of the binary in each frame is calculated using the difference flux from the
subtracted image and a flux measured on the unsubtracted reference image.

Since accurate eclipse depths are critical in the characterization of an
eclipsing binary, we checked the ISIS eclipse depths by measuring the eclipse
depth for a primary eclipse on a single night using a traditional
curve-of-growth analysis \citep{stetson1990}.  We used the ${\scriptstyle
  MATCH}$ and ${\scriptstyle MASTER}$ tasks to remove zero-point differences
between images introduced by variations in atmospheric transparency, airmass,
and exposure time. We found that the derived eclipse depth from ISIS and the
curve of growth analysis were identical within the uncertainties.

Figure \ref{a259lcfull} shows our full set of ground-based photometric
observations phased to the orbital period, and Figure \ref{a259lcecl} is
zoomed in on the primary and secondary eclipses,  showing the observations
near eclipse minima that were used to model the binary.

\subsection{Spectra}\label{spec}

Spectroscopic observations were obtained as part of the WIYN Open Cluster
Study \citep[WOCS;][]{mathieu}.  We use the observations of \citet{hole}, who
presented high-precision RVs for 1207 stars in NGC 6819, and more recent WOCS
spectroscopic observations of \citet{milliman}.  A detailed explanation of the 
acquisition and reduction of the spectra, and the RV measurement uncertainties 
(typically, $\sigma=0.5\kms$ for single stars) of WOCS observations can be 
found in \citet{geller}; below we provide a brief description.

WOCS uses the WIYN\footnote{The WIYN Observatory is a joint facility of
  the University of Wisconsin-Madison, Indiana University, Yale University and
  the National Optical Astronomy Observatory.} 3.5-m telescope on Kitt Peak
and the Hydra multi-object spectrograph (MOS) instrument, which is a fiber-fed
spectrograph capable of obtaining about 80 spectra simultaneously (usually 10
fibers to measure the sky, and about 70 fibers observing stars).
We observed NGC 6819 over 35 separate observing runs using 
the echelle grating, which provides a spectral resolution of 
$\sim 15 \kms$.  The spectra are centered at 513 nm with 
a 25 nm range in order to cover narrow absorption lines 
around the Mg \Rmnum{1} b triplet.  Spectroscopic 
observations of WOCS 40007 were completed using 1- or 2-hour 
integrations per visit that were split into three  separate 
integrations to allow for the rejection of cosmic rays.

Standard spectroscopic image processing was also done within {\it IRAF}.  
Calibration of the spectra from each epoch used one associated 200 s flat
field and two 300 s ThAr emission lamp spectra bracketing science
integrations.  RVs for WOCS 40007 were extracted using TwO Dimensional
CORelation \citep[TODCOR;][]{zucker} against an observed solar template
spectrum.  Additionally, we correct for fiber-to-fiber radial-velocity offsets
present in the Hydra MOS, as described in \citet{geller}. The offsets ranged
from $-0.70$ to $0.37 \kms$, with an rms value of $0.16 \kms$.

For the purposes of this paper, we have not attempted to estimate the impact
of lines moving into or out of the spectroscopic window. \citet{meibom}
examined this effect for a binary in NGC 188 for the exact same instrument and
setup, finding that it had an effect of up to 3 \kms ~ near primary eclipse
and the systematic error was phase dependent. However, at most phases the
systematic errors were less than 1 \kms.  Figure \ref{a259rv} shows our RV
observations phased over the binary orbit, along with the velocity residuals
(observed minus computed). In addition to the three observations we omitted
due to potential effects of eclipses, there are three measurement epochs that
show significant deviations from the model. We will address the issue of
spectral window corrections in a future paper.

\subsection{Binary Component Magnitudes and Temperatures}\label{decomp}

Before modeling the spectroscopic and photometric data, we first derive
constraints on the temperatures of the individual stars. 

To derive spectroscopic temperatures, we first selected
the 22 spectra where the primary and secondary are well separated in RV space,
as this reduces complications involving blended lines.  We selected templates
from a library of ATLAS9 synthetic spectra \citep{ck03,ck04} calculated by Jon
Morse for an extensive grid of model atmospheres computed using the ATLAS9
code developed by Kurucz, originally for the CfA Digital Speedometers
\citep[e.g.][]{latham} and broadened to the WIYN spectral resolution. The
templates were chosen to have $\log({\rm g})=4.0$ and $T_{\rm eff}$ in a range
from 3500 K to 7000 K (in steps of 250 K). Non-rotating templates were chosen
because we were unable to identify significant rotational broadening
in the spectra. For each
observation we cross-correlated all combinations of these templates against
the observed spectrum using TODCOR, and determined the template combination
that returned the highest 2D correlation peak height.
This analysis returned temperatures of $6350\pm150$ K for the primary
component and $5930\pm150$ K for the secondary. The standard 
error in the mean from the 22 spectra was $80$ K, but we have increased the uncertainty to $150$ K
to account for the possibility of a systematic offset in temperature when 
correlating observed spectra against synthetic templates 
\citep[e.g.][]{meibom}.

We also have standard-system photometry of WOCS 40007 in $B,~V,$ and $I_c$
from \citet{talamantes} that can be used to derive photometric
temperatures. ($R_c$ was not calibrated.)  The calibrated observations were
taken on a single night, which unfortunately was the night of a primary
eclipse.  Three to four observations per filter were taken outside of eclipse, 
but these may still be systematically faint by approximately 0.01 mag because 
they were not at quadrature when the system light is maximum.

The discovery of a faint tertiary star (see \S \ref{tert}) adds a complication
to the disentanglement of the light of the stars, but because the secondary
eclipse is total, the photometry of the secondary star is well constrained.
To derive the photometry of the primary star, we need to make the assumptions
about the photometric properties of the tertiary star. At present, the
tertiary's photometry is weakly constrained by the assumption that it falls on
the cluster fiducial line in the CMD, and by the minimum mass derived in \S
\ref{tert} (which sets a lower limit on the brightness of the tertiary).  The
need to have the primary star fall on the cluster fiducial line puts an
effective upper limit on how bright the tertiary can be.
The measured secondary eclipse depths ($\Delta m_s$) and corresponding
magnitudes of each component are listed in Table \ref{tablecomponents}.  
We find the uncertainties for the secondary eclipse depth by determining the
interquartile range of measurements during totality and propagate this through
the calculations to determine the magnitude uncertainties.

As seen in Figure \ref{figcomponents} (c) and (e), the
photometry of the primary component of WOCS 40007 is consistent with the main
sequence line in both $B-V$ and $V-I_c$ for an appropriate choice of
tertiary in the CMD, but the secondary component is only consistent
with the main sequence in $B-V$. In $V-I_c$, the secondary is redder
than the main sequence by $\sim 0.06$ mag. Since the
secondary eclipse is total, the properties of the secondary component
are independent of the assumptions about the tertiary.  This is true 
as long as the eclipse depths and system photometry are measured correctly.
Because this shift in $V-I_c$ affects the determination of the temperature 
of the secondary star, we delve a little into possible causes. We first 
compared our photometry to that of previous studies. Only \citet{rosvick} 
and \citet{hole} have previously presented $I_c$ photometry, and in both 
cases, WOCS 40007 is shifted similarly to the red {\it relative to the main 
sequence in the same dataset}.

Using the secondary eclipse depths to decompose the system photometry, we
still find that the secondary star is redder than the main sequence at the
same brightness level. Looking at $B$ and $V$ photometry from these sources 
as well as \citet{kalirai}, the position of the secondary star in the CMD is 
much more consistent with the main sequence (see Fig. \ref{figcomponents}).

Because WOCS 40007 contains a relatively short period binary, we considered
the possibility of photometric variations due to spots and stellar activity.
We detect non-synchronous spot modulation at about the 1\% level outside of
eclipse in \kep photometry, but due to systematic trends introduced by \kep, 
longer timescale variations are harder to study reliably.  NGC 6819
was observed at different epochs in previously published datasets, so that
longer term photometric variations might be identifiable if large enough.  To
produce clearer comparisons with our photometry, we corrected the respective
datasets for zero-point differences determined from all stars in common.
There are often significant zeropoint offsets between datasets (see Table
\ref{zpts}), but there is only evidence of problematic color-dependent
trends in the $B$ photometry of \citeauthor{hole}. The results from the
decomposition of the photometry from each source are given in Table
\ref{tablecomponents}. The differences cover a range of up to 0.07 mag (in
$B$) in brightness and 0.05 (in $B-V$) in color and are
similar to twice the semi-interquartile ranges in the photometry comparisons
(Table \ref{zpts}). 

At present we do not have a satisfying explanation for the unusual $V-I_c$
color of the secondary star. Because of potential issues with the $I_c$
photometry, we will focus on the isochrone fit in the $(V,B-V)$ CMD (see
Section \ref{seccmdiso}). In order to constrain the temperature of each
component, we adopt the metallicity [Fe/H]$= +0.09\pm0.03$ from
\citet{bragaglia} and use our value for the cluster reddening (see
\S \ref{reddist})
and the empirical temperature scale of \citet{casagrande}.
\citet{vandenberg2010} employed this temperature scale to transform their
theoretical models to the color-magnitude plane, and found that it produced
excellent agreement with the photometry of stars similar to NGC 6819's
upper MS stars in well-measured open clusters with solar (M67) and
super-solar (Hyades) abundances bracketing NGC 6819's composition.  For the
final temperature determination, we employ the photometry from \citet{kalirai}
since it possesses the highest precision in $B$ and $V$ photometry, as judged
from the smaller scatter among their main sequence stars (see Figure
\ref{figcomponents}).  We find, and adopt, temperatures for the primary and
secondary star of $6240\pm80$ K and $5950\pm70$ K, respectively.  These
photometric temperatures are given in the last column of Table
\ref{tablecomponents} and are consistent with the temperatures found 
spectroscopically ($6350\pm150$ K; $5930\pm150$ K).

\subsection{Binary-Star Modeling}\label{elcmod}

Before computing binary star models, we removed some data points from
consideration because of the possibility they could bias the results. Most
importantly, we chose to restrict the photometric dataset to observations in
and near the eclipses. Because the binary is well separated and not observed
to vary significantly outside of eclipse, the out-of-eclipse observations
contain little physical information on the binary. Variations in the
photometry out of eclipse tended to make it more difficult to find the best
fit to the eclipses (see Figure  \ref{a259lcecl}). Additionally, we shifted the
$V$ secondary-eclipse data from 2010 September 8 ($\phi=0.46-0.52$) by +0.062
magnitudes.  Initial light curves of 2010 September 8 showed the observations
to be consistently fainter than \citet{talamantes} observations over the same
phase range. The observed difference is similar in amplitude to variations for
rotating stars in young clusters (e.g. \citealt{meibom11}), but is smaller
than the short-term variations ($\sim 0.01-0.02$ mag) seen in \kep
observations of this system. A smaller number of $R_c$ observations were
obtained on 2010 September 8, but no noticeable shift was evident in comparing
to $R_c$ observations on other nights. In the absence of a clear explanation,
we opted to simply apply a zeropoint correction to the $V$ data.  We
determined the shift by comparing out-of-eclipse observations from 2010
September 8 with those of 2008 June 5. Finally, we eliminated a handful of
radial-velocity observations that were taken close to eclipse phases because
they contributed disproportionately to the overall $\chi^2$ of the fits and
may have been affected by line blending.  The eliminated points are shown as
squares in Figure \ref{a259rv}.

Models of the photometric and RV data were computed using the ELC code
\citep{orosz}.  We employ ELC's genetic and Markov optimizers to arrive at the
best-fit solution. ELC employs the $\chi^2$ statistic to describe the
goodness-of-fit.  The total $\chi^2$ is the sum of individual $\chi^2$
contributions from photometry for each filter, radial velocities for each
component, and additional contraints derived from other observations.
For WOCS 40007, our baseline model has eight free parameters: the inclination $i$;
velocity semi-amplitude of the primary $K_1$; temperature of the primary
$T_{\rm 1}$; temperature ratio $T_{\rm 2}/T_{\rm 1}$, ratio of the primary
star radius to the semi-major axis $\rr_{\rm 1}/a$; ratio of the radii
$\rr_{\rm 1}/ \rr_{\rm 2}$; orbital period $P$; and the time of primary
eclipse minimum $t_0$.  To examine the effects of the limb-darkening
treatment, we computed two kinds of models. First, we used a quadratic
limb-darkening law, assuming one coefficient for each star (from
\citealt{claret}) and fitting for the other. This was done in the hope that an
error in the assumed coefficient will be compensated for by a change in the
value of the other because the coefficients are generally correlated
\citep{southworth}. Second, we employed PHOENIX model atmospheres
\citep{hauschildt} to describe the variation of emitted intensity with
emergent angle, which removes the need to assume limb-darkening coefficients.
We include both limb-darkening and model atmosphere results in Table
\ref{tabsolutions}.  

A cursory examination of the light and RV curves for WOCS 40007 reveals that
there is little or no eccentricity.  In test runs that allowed the
eccentricity to vary, we found a best-fit solution with eccentricity
$e=0.0003\pm0.0001$ and argument of periastron passage
$\omega=90.0\pm0.1\degr$.  Therefore, we set the eccentricity to zero for all
other model fits, which makes it possible to tightly constrain the orbit
parameters $K_1$ and $q=\mm_2/\mm_1$ entirely using RV observations. We split
the modeling into separate runs for the photometric and spectroscopic data.
Eclipse timing variations from \kep observations indicate the presence of a
tertiary component (see \S \ref{tert}), which meant that $K_1$ and $q$ would
be determined most precisely after the tertiary star orbit is fit and
corrected for. Therefore, in our final fits of the photometry, we set the
values of $K_1$ and $q$ as observed constraints (described below).

Following the completion of the photometry model runs, we used the best-fit
parameters (particularly $i$, $P$, and $t_0$)
to determine the primary and secondary masses
in a final RV model run.  By completing the RV and photometric model
optimization runs for WOCS 40007 separately, the relative weights of the RV
and photometric measurements are not a concern, as they would be in a
simultaneous fit to both.

To ensure that we derive realistic parameter uncertainties, we scale
the error bars for our observations in order to obtain a reduced 
$\chi_\nu^2 = \chi^2_{\rm phot}/ \nu \approx 1$ (where $\nu$ is the number of 
degrees of freedom) for photometry in each filter and for the velocities measured for each star.
Because the measurement uncertainties generally need to be scaled
upward, this results in conservative uncertainties on the binary star
model parameters because it allows a greater range of models to fit
the data.  The scatter around the best-fit model for the radial
velocities gives an alternate estimate of the measurement uncertainty:
3.14$\kms$ for the primary star and 3.34$\kms$ for the secondary star.  The
scatter is significantly larger than the measurement precision for single stars
\citep[0.4 km s$^{-1}$;]{hole} due to spectral line blending, and with further
correction to the RVs derived from our TODCOR analysis (see \S \ref{spec}) is
anticipated to decrease.  The typical scaled uncertainty in each photometric
observation is $\sim 0.015$ magnitudes for $B,~R_c,$ and $I_c$, and $\sim
0.035$ magnitudes for $V$.
Uncertainties for each free and derived parameter were 
determined by projecting the $n$-dimensional $\chi^2$ 
function onto each parameter.
$1\sigma$ confidence intervals for the parameters are set where the 
lower envelope of $\chi^2$ values reaches $\chi^2_{min} + 1$ \citep{avni}.
Figure \ref{a259chi} shows the plots of parameter versus $\Delta\chi^2$.

Our best fit values for the free and derived parameters of the
eclipsing binary can be seen in Table \ref{tabsolutions}.  The top
portion of the table lists parameters that were fit during ELC runs,
while the lower portion of the table lists parameters that were
calculated from the values derived by ELC.  For completeness, we
include both model-atmosphere and limb-darkening results in the table. Both
alternatives (use of model atmospheres or tabulated limb-darkening
coefficients) will result in a degree of systematic error due to
systematic errors in selecting the coefficients, or due to inappropriate
vertical structure of the model atmospheres. 
Indeed, some parameters in Table \ref{tabsolutions} are not consistent within
1-$\sigma$ uncertainties (e.g., $\rr_1$ and $\rr_2$). After some
investigation, we found that this was because the atmosphere models were
allowed to be nonspherical. When the limb darkening models were also allowed
to be non-spherical, the best fit model agreed very well with the model
atmosphere results. In spite of this, we adopt the tabulated limb-darkening
results (with the assumption of sphericity) in later analyses. Our reason is
that the derived radii can be sensitive to light curve shape near first and
fourth contact, and star spots can affect the curves in these critical parts.
If a spot is roughly centered on the side facing the other star, it can
produce increasing brightness after the eclipse has completed, which is
similar to the effects of nonsphericity. There is clear evidence of
nonsynchronously rotating spots in \kep light curves that can produce these
effects, but no evidence of other out-of-eclipse light variations that would
be associated with non sphericity.

Fits that are more
constrained in this way understandably produce greater deviations
between observations and model near eclipse ingress and egress, and therefore
result in some disagreement in the radius-related parameters given in
Table \ref{tabsolutions}. By allowing some of the limb-darkening
coefficients to be free parameters, we hope to minimize systematic
errors in the radii of the stars.

\begin{figure}
\begin{center}
\includegraphics[width = \textwidth]{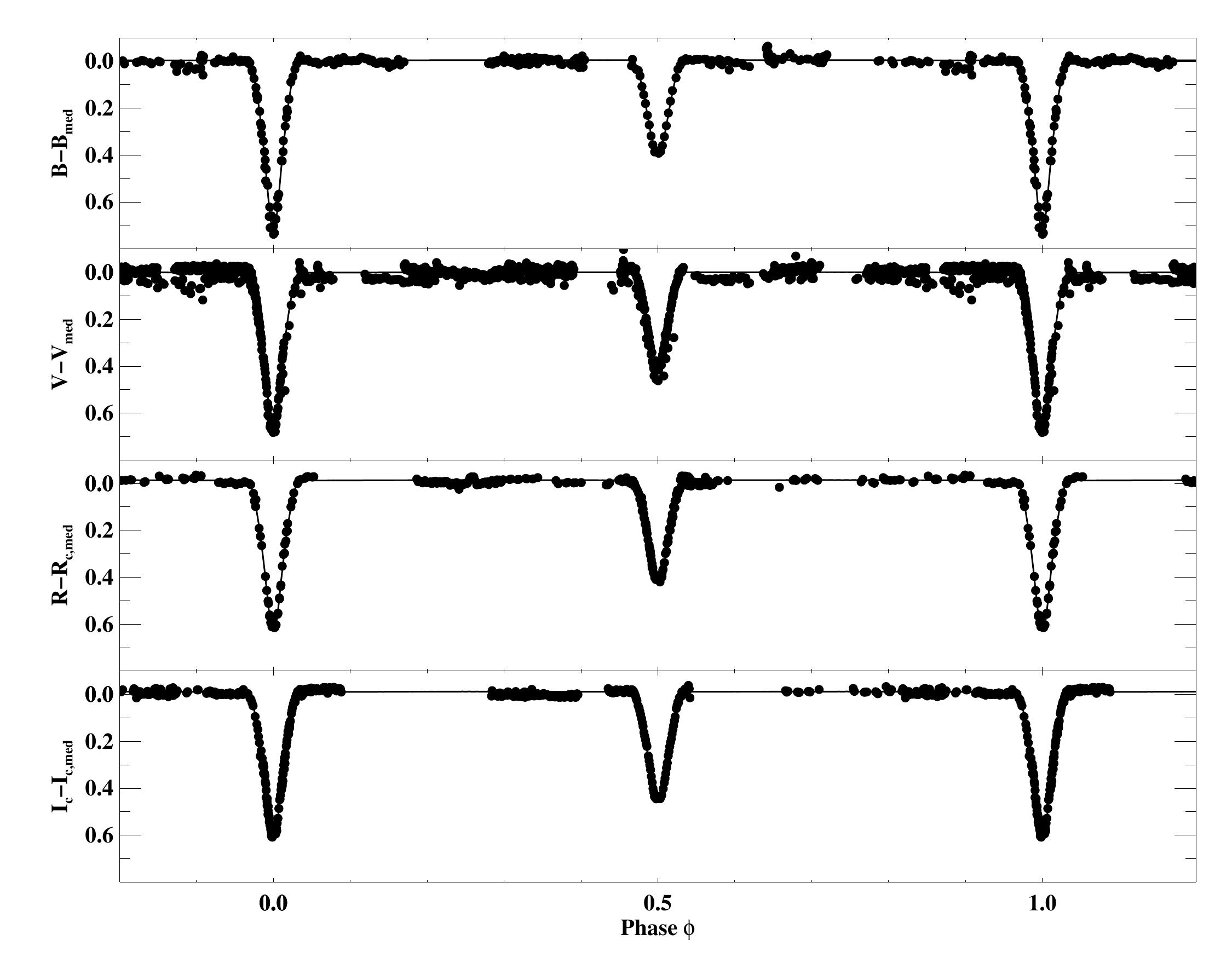}
\caption{Photometry for WOCS 40007, phased to the orbit period for each
  filter. The solid line is the ELC model fit. }
\label{a259lcfull}
\end{center}
\end{figure}

\begin{figure}
\begin{center}
\includegraphics[width = \textwidth]{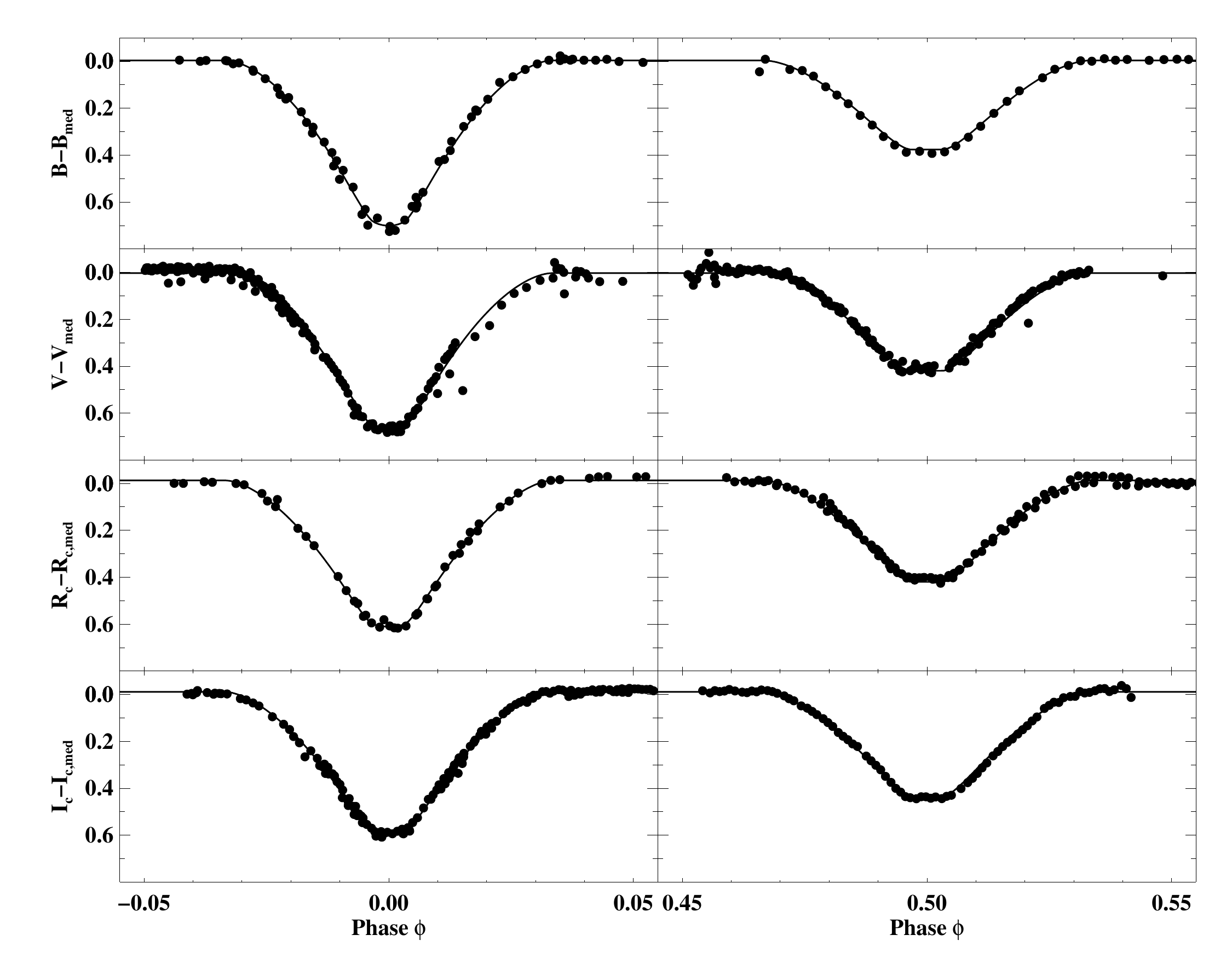}
\caption{Photometry for WOCS 40007, zoomed in on the primary (left column) and
  secondary (right column) eclipses. The black line corresponds to the ELC
  model fit.  This restricted set of data was used to determine the best fit
  model.  }
\label{a259lcecl}
\end{center}
\end{figure}

\begin{figure}
\begin{center}
\includegraphics[width = \textwidth]{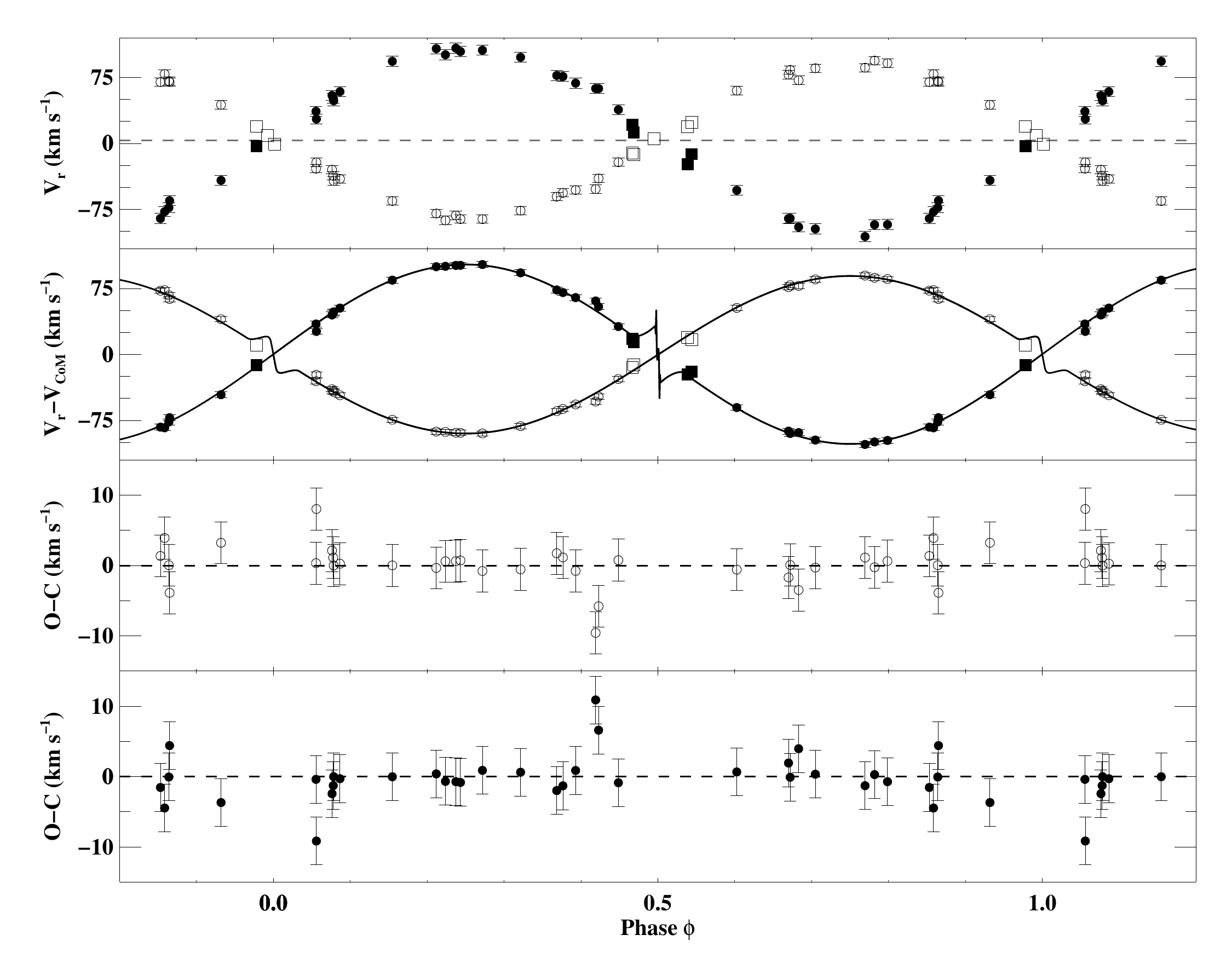}
\caption{Radial-velocity observations for WOCS 40007 phased to the orbital
  period. The top panel shows the uncorrected RV data, while the second panel
  shows the RV data with the center-of-mass velocity variation due to the
  tertiary component removed.  The black line in the second panel corresponds
  to the ELC model.  The third and fourth panels show the $O-C$
  diagrams for the eclipsing binary components. In all
  cases, open circles are primary star observations and filled circles are
  secondary.  Open and filled squares are primary and secondary star
  observations that were not included in the ELC fit due to blending of the
  stellar spectra.
\label{a259rv}}
\end{center}
\end{figure}

\begin{figure}
\begin{center}
\includegraphics[width=\textwidth]{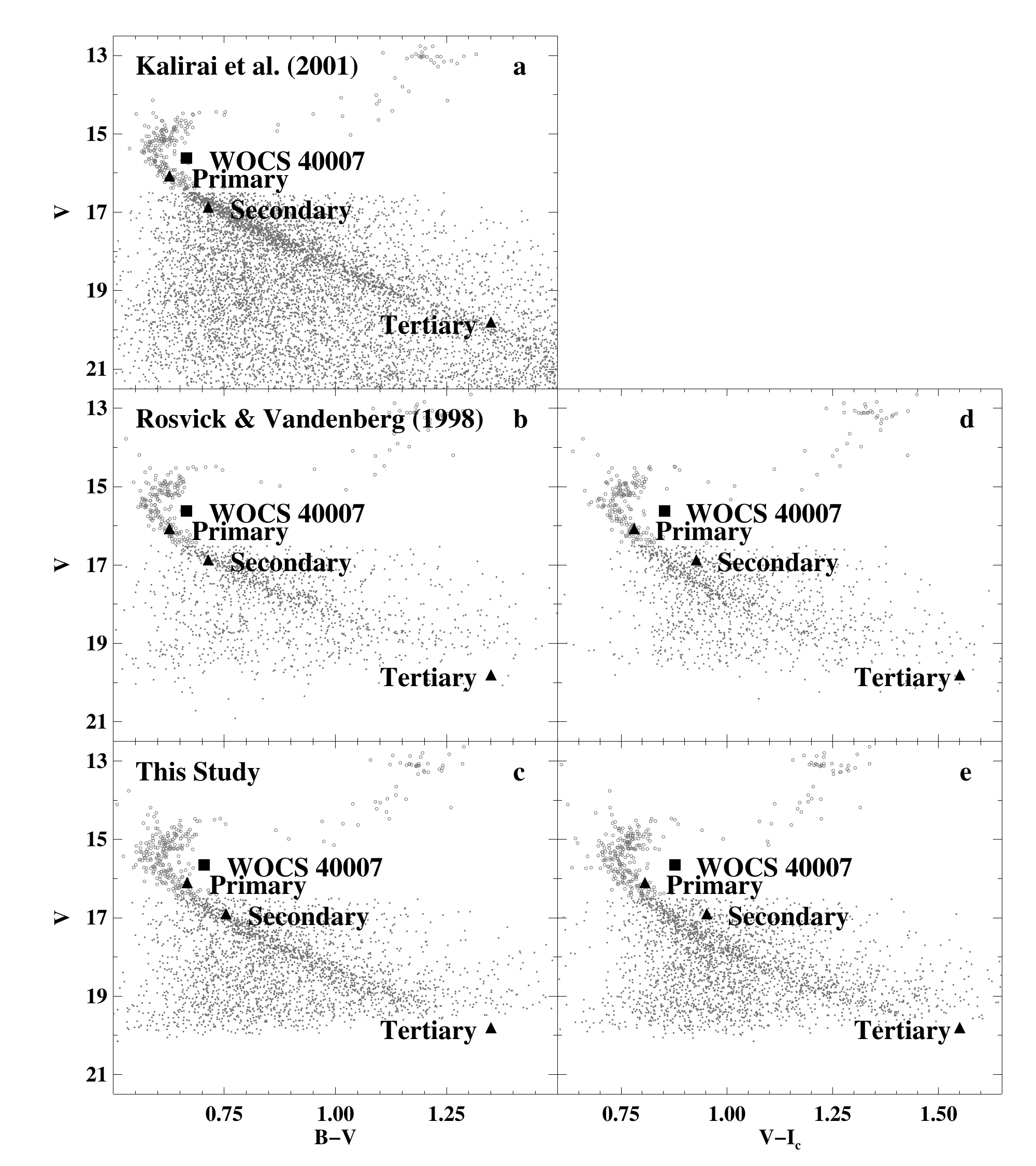}
\caption{ Derived locations of the DEB components in the CMD of NGC 6819 using
  the calibrated photometry of \citet[top panel]{kalirai}, \citet[middle
    panel]{rosvick} and this study (bottom panel).  The system photometry
  (squares), and that of the stellar components (triangles) are shown. Known
  cluster members are plotted as open circles for magnitudes brighter than the
  WOCS faint limit ($\lesssim 16.5$ mag).  }
\label{figcomponents}
\end{center}
\end{figure}

\begin{figure}
\begin{center}
\includegraphics[width = \textwidth]{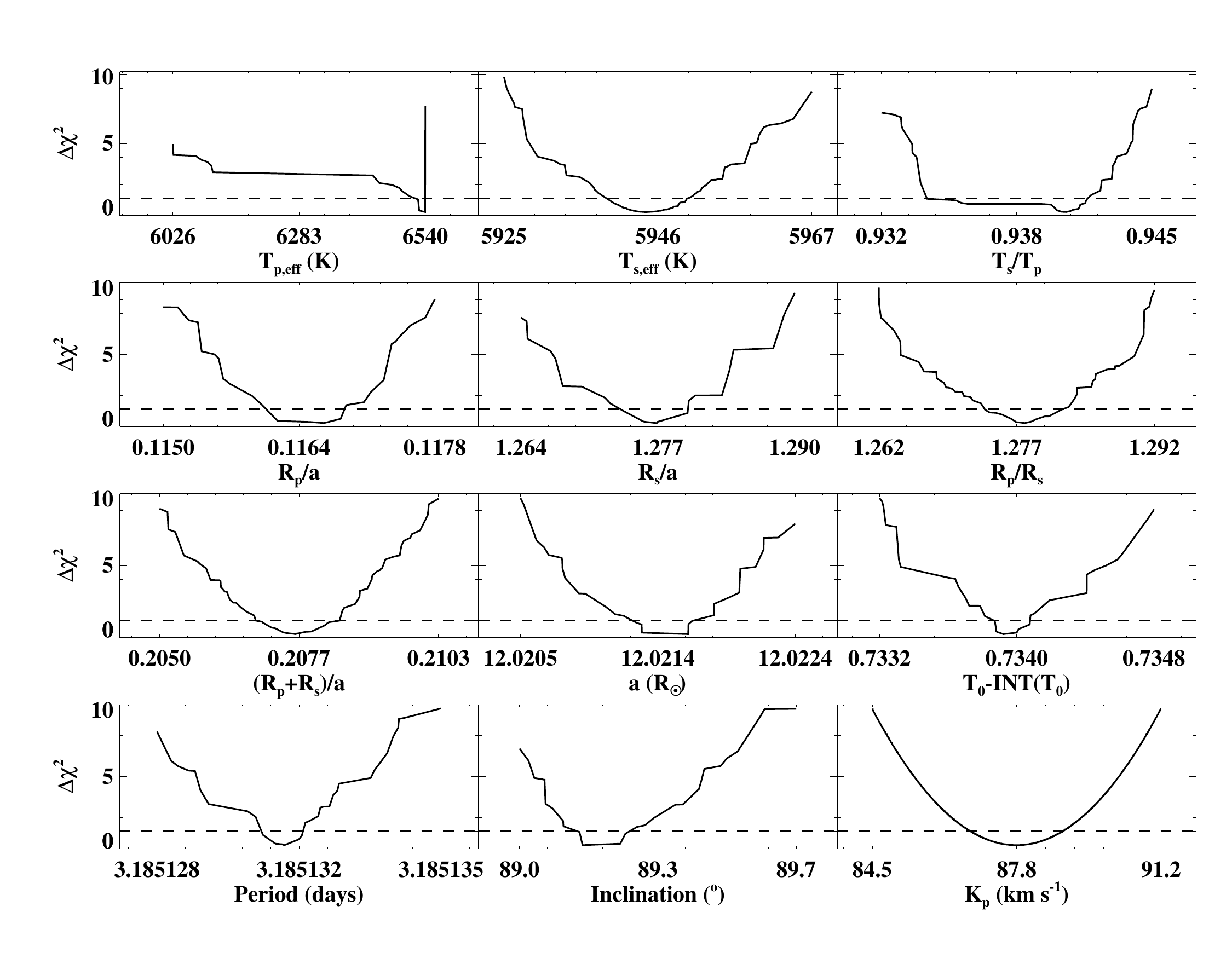}
\caption{ $\Delta\chi^2$ vs. fitted parameters for WOCS 40007.  The solid 
line is the lower envelope of all models. The horizontal dashed line 
corresponds to $\Delta\chi^2=1$, representing the $1\sigma$ uncertainty in 
each parameter.
}
\label{a259chi}
\end{center}
\end{figure}

\begin{figure}
\begin{center}
\includegraphics[width = \textwidth]{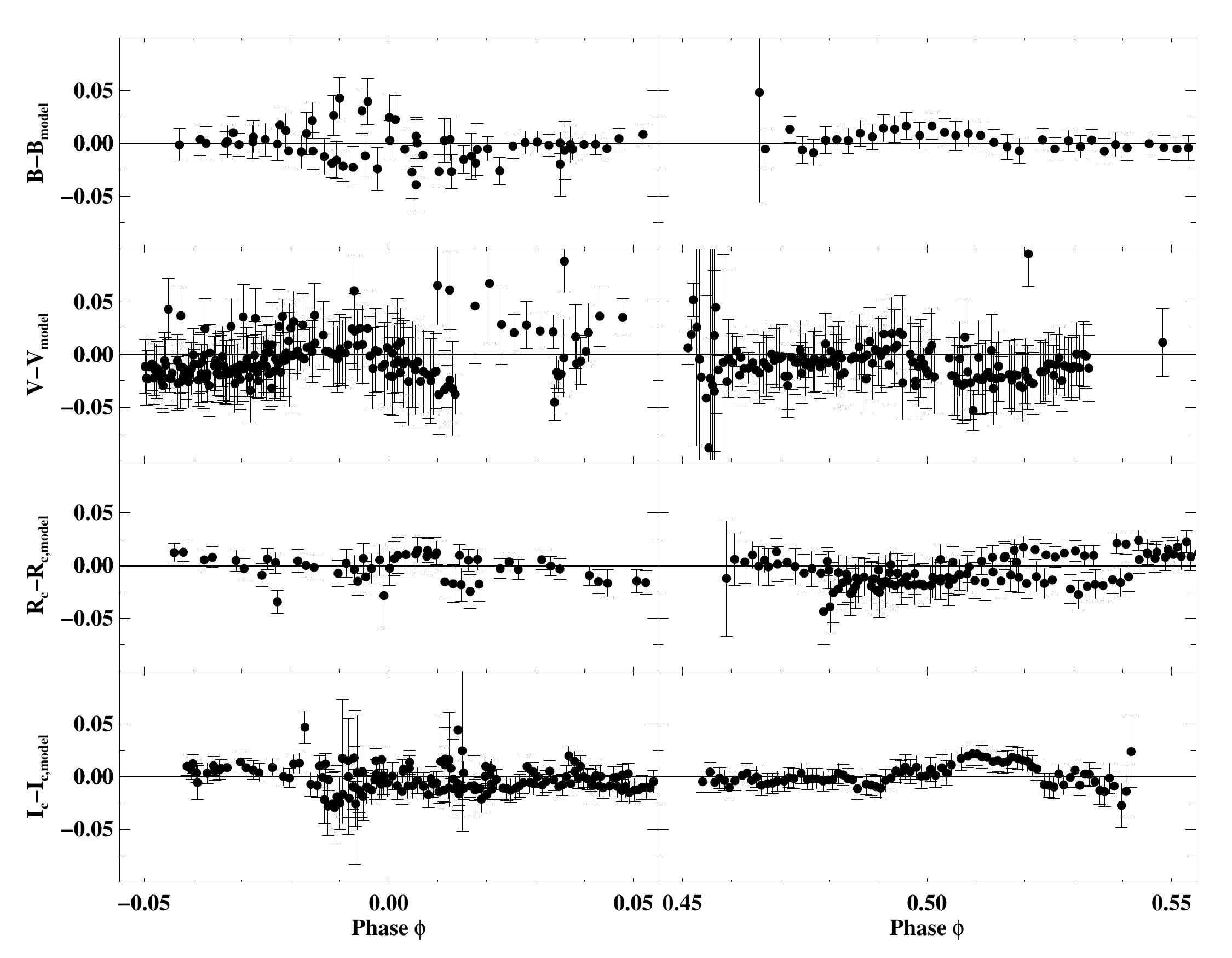}
\caption{
$O-C$ diagrams corresponding to the photometric solution 
presented in Figures \ref{a259lcfull} and \ref{a259lcecl}, 
zoomed in on the primary (left column) and secondary (right 
column) eclipses.  Data points included in this figure 
correspond to the data used by ELC to determine the best 
fit model.
}
\label{a259lcoc}
\end{center}
\end{figure}

Figures \ref{a259lcfull} and \ref{a259lcecl} show the ELC 
model over-plotted on our $B,~V,~R_c$, and $I_c$ photometry 
and Figure \ref{a259lcoc} displays the $O-C$ diagrams for the 
four light curves.  Our RV fit for WOCS 40007 is presented in Figure 
\ref{a259rv}.  Error bars are not shown in Figures 
\ref{a259lcfull} and \ref{a259lcecl} in order to clearly show 
the behavior of the observed light curve compared to the ELC 
model.  
Our photometric observations of WOCS 40007 clearly show that the secondary
eclipse ($\phi = 0.5$) is total, allowing precise measurements of
the radii of the primary and secondary components.

\subsection{Detection of a Tertiary Star}\label{tert}

As seen in Figure \ref{figcomponents}, WOCS 40007 is much redder than expected
in the $(V,V-I_c)$ CMD.  This hinted that a low-mass star could be blended
with the eclipsing binary, though the presence of a third star has not been
identified in the spectra.  We have two sets of long-term
measurements that could reveal such a star if it orbits the eclipsing binary:
the center-of-mass velocities of the eclipsing binary monitored since 1998,
and nearly continuous eclipse timing by the \kep mission since 2009 and by
earlier ground-based observers on a much more limited basis.

We did not attempt to model \kep photometry here because it is difficult to be
completely sure that instrumental trends are entirely removed, and that
contamination from other stars is fully corrected for --- this will be done in
a later paper. As we are only interested in eclipse timing, we will not delve
deeply into the reduction.  The eclipses of WOCS 40007 are very clearly
visible in the raw pixel data, although we did detrend the light curves for
instrumental effects before determining eclipse timings using the method of
\citet{kwee}. We have also corrected for a 66.184 s error in the \kep data,
where times had been reported in UTC rather than TDB (barycentric dynamical
time).  Here we used long-cadence data from quarters 1-13 (excluding quarters
6 and 10 because the cluster fell on a dead CCD module), and the barycentric
Julian dates (BJD) of middle eclipse typically had uncertainties of
approximately 15 s.  The eclipse timings are given in Table \ref{etiming}. For
ground-based observations, the heliocentric Julian dates were converted to BJD
using an online calculator\footnote{\tt
  http://astroutils.astronomy.ohio-state.edu/time/hjd2bjd.html}
\citep{eastman}. Initially, the \kep observations appeared to be consistent
with a constant period, but by quarter 9 there was a noticeable deviation from
a linear trend in the observed minus computed ($O-C$) diagram.  A much earlier
ground-based eclipse observation from \citet{talamantes} also cannot be
explained by a simple linear ephemeris. The gravitational influence of a third
orbiting body can produce deviations from a linear ephemeris due to light
travel time variations.

There would also be significant variations in the center-of-mass velocity of
the binary introduced by a third body, so we did a simultaneous three-body fit
to both the eclipse timings and the radial velocities of the primary and
secondary stars, assuming that the third body is on a Keplerian orbit and
interacting with the barycenter of the eclipsing binary. We therefore fit for
the following parameters: the periods of the eclipsing binary $P_b$ and
tertiary $P_3$; the times of primary eclipse for the binary $t_b$ and time
of periastron for the tertiary $t_3$ (in BJD); and the eccentricity $e_3$ and
argument of periastron $\omega_3$ of the tertiary orbit (the eclipsing binary
was assumed to be circular). These values are quoted in the top half of Table
\ref{terttab}. The largest effects on eclipse timing happen near periastron,
which fortuitously occurred while \kep has been observing. Our earliest
eclipse timing (from ground-based observations in July 2001) appears to have
occurred about the time of the previous periastron passage.  This observation
only phases with recent photometry if the binary period (in the absence of
light travel time effects) is significantly smaller than is being observed now.

\begin{figure}
\begin{center}
\includegraphics[width = \textwidth]{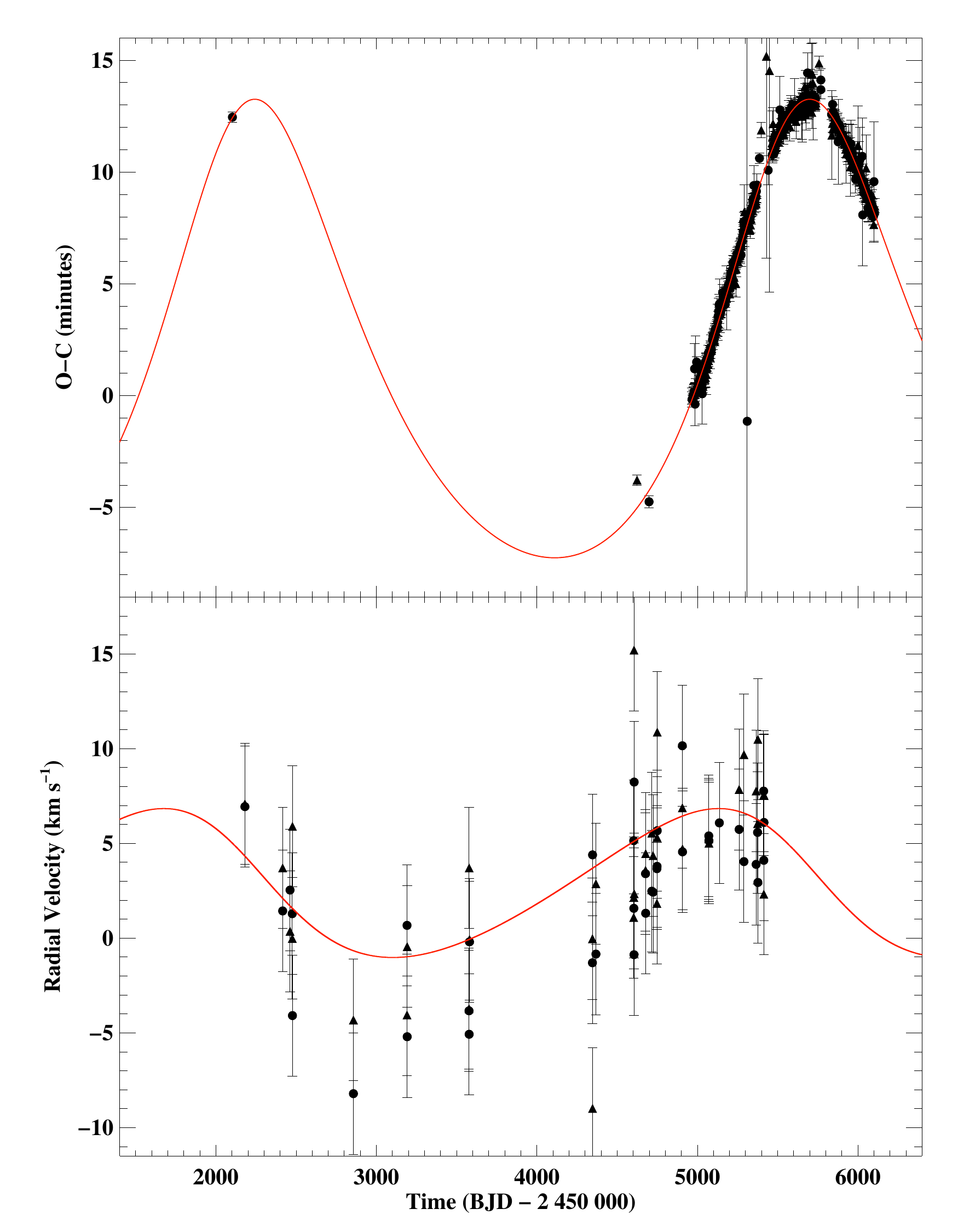}
\caption{ Observed minus computed values for primary ({\it solid dot}) and
  secondary ({\it solid triangle}) eclipse timing ({\it top panel}) and radial
  velocities of the barycenter as derived from the primary and secondary stars
  ({\it bottom panel}). The curves show the best fit from three-body models.
 The outliers in the eclipse timing panel are ground-based observations. }
\label{tertorbit}
\end{center}
\end{figure}

We computed radial velocities from the model, and compared with the 
observed radial velocities  (see Figure \ref{tertorbit}).  The reduced 
$\chi^2$ values for the best
unconstrained fit and the eclipse timing model are quite similar (0.99
vs. 1.19). The velocity semi-amplitudes of the binary components $K_1$ and
$K_2$, the binary barycenter $K_b$, and system velocity $\gamma$ given in
Table \ref{terttab} were derived using the observed radial velocities and the
eclipse timing model for the tertiary orbit.  The radial velocities and
eclipse timings both appear to have covered approximately one orbital period
for the tertiary. We did not see evidence of an eclipse of the tertiary by
either star in the binary in \kep photometry to date, so the inclination of
the tertiary orbit is undetermined. Future
\kep eclipse timing and ground-based radial velocities will help nail down
the tertiary period and velocity amplitude. However, the solution of the
tertiary orbit does not affect the measurements of the masses and radii of the
eclipsing binary components significantly.

With the total mass of the eclipsing binary known (as discussed
below), the velocity variation sets a lower limit for the mass of the
tertiary. This would be the actual
mass if the tertiary orbit has inclination $i=90\degr$. Using models
of the radial velocities, we find that the minimum mass is about $0.56
\msun$.

Based on the lower limit to the tertiary mass, we can determine
a faint limit for the tertiary star from stellar evolution models. We
estimate the tertiary photometry by trying possible MS magnitudes
between 19.0 and 20.5 in $V$. On the faint end, the tertiary
properties are constrained by restricting the star to be on the
cluster main sequence with a mass above the minimum derived above. On
the bright end, the tertiary is constrained by its effects on the
derived properties of the primary. The primary magnitude is dependent
on the choice of the tertiary magnitude, while the secondary component
remains at a fixed magnitude because its properties are determined
from the secondary eclipse depth, which provides information on the
combined primary and tertiary light (see Section \ref{decomp}).
Selecting the best photometric match, we adopt a tertiary component
with a $B$ magnitude of $21.5\pm0.4$, $V$ magnitude of $20.1\pm0.4$,
$R_c$ of $19.3\pm0.4$, and $I_c$ magnitude of $18.5\pm0.4$.

The motion of the eclipsing binary induced by the tertiary has some small
effects on the modeling of the binary star data, and these were noted earlier
in \S \ref{elcmod}. However, it should be noted that the period of the
eclipsing binary $P_b$, the velocity semi-amplitudes $K_1$ and $K_2$, and the
mass ratio $q$ given in Table \ref{terttab} are the closest things to ``true''
--- the values that would be observed in the absence of the tertiary.

\section{Photometric and Spectroscopic Results}\label{obsresults}

\subsection{Ephemeris for WOCS 40007}

Because the gravity of the tertiary star causes shifts in the position
of the eclipsing binary system relative to the observer, a linear
ephemeris will be inadequate to predict eclipse timings to better than
a few minutes.
Readers should note that most of our ground-based observations were taken
during times when the eclipsing binary was in a part of its orbit moving away
from Earth with a nearly constant speed, and the extra light-travel time added
during each orbit cycle resulted in a larger period measured at Earth (see
Table \ref{tabsolutions}). However, this is not the ephemeris that would be
determined from observations while moving along with the binary. As stated
earlier, our fit for the true orbital period of the eclipsing binary is given
in Table \ref{terttab}.

\subsection{Cluster Membership}

Previous information about the membership of WOCS 40007 was rather
limited.  The system was too faint to be in the proper-motion study of
\citet{san72}, and was identified by \citet{hole} as a binary with unknown
membership because its system velocity could not be reliably
determined at the time. We are in a position now to discuss the binary's
membership using three different criteria.

First, based on its position $3\farcm37$ from the cluster center (1.4 core
radii; \citealt{kalirai}) and on the relative density of cluster and field
stars at its magnitude level as estimated from Figure 3 of \citeauthor{kalirai},
we determine a membership probability of 53\%. This is the weakest constraint
on membership though. Second, using the systemic velocity measured from our
spectroscopic solution, we can calculate the membership probability using
results of the radial-velocity survey of \citet{hole}:
  \[ p(v)=\frac{F_c(v)}{F_f(v)+F_c(v)} \]
where $F_c(v)$ and $F_f(v)$ are one-dimensional Gaussian functions describing
the cluster and field RV distributions, respectively.  We employed the
Gaussian fit parameters given in \citet{hole}, and find that WOCS 40007 is a
cluster member with a 84\% probability.

The third membership measure involves proper-motion measurements
(Platais et al., in preparation) combining old photographic plates
with CCD observations using the 3.6~m Canada-France-Hawaii Telescope
(CFHT). Although the star is isolated and optimally exposed on the
images, there were somewhat larger than average astrometric errors in
declination (0.27 mas y$^{-1}$, versus 0.14 mas y$^{-1}$ in right
ascension) likely due to coma present on the Hale 5 m telescope
plates, which were the backbone of the astrometry. Despite this 
complication, we determine the astrometric membership probability to be 
97\%.  As a result, we judge WOCS 40007 to be a cluster member.

\subsection{Distance Modulus}\label{secdm}

The absolute magnitude and apparent distance modulus can be calculated using
the radius from the binary star analysis and a temperature estimate.  We
calculated bolometric corrections in $V$ from \citet{vandc}, and used them
along with the photometric temperature estimates to find $(m-M)_V =
12.49\pm0.10$ and $12.39\pm0.11$ for the two components of WOCS 40007. The
distance moduli for the stars agree within the uncertainties, and the averaged
distance modulus from the binary star components [$(m-M)_V=12.44\pm0.07$] is
slightly larger than found from fits to older isochrone sets (12.35,
\citealt{rosvick}) or from the red clump analysis in \S \ref{reddist}.  The
largest contributor to the uncertainty in this distance modulus by far is the
temperature. If we use the spectroscopic temperatures or the radii from the
model atmosphere runs instead, the derived distance moduli only differ by
about 0.03 mag and there is better consistency. However, with the other
uncertainties that are present (e.g., the photometry of the primary star) this
is not a clear reason to prefer one set of values over another.

\section{Comparison with Stellar Evolution Isochrones}\label{iso}

We use four different sets of theoretical stellar evolution isochrones to
interpret the masses, radii and photometry of the WOCS 40007 stars.  The four
isochrone sets are: BaSTI \citep{pietrinferni}, Dartmouth Stellar Evolution
Program \citep[DSEP;][]{dotter}, Victoria-Regina \citep[VR;][]{vandenberg2006}
and Yonsei-Yale \citep[Y$^2$;][]{demarque}.  We generate scaled-solar
isochrones with parameters closest to the spectroscopic abundance for NGC 6819
([Fe/H]$=+0.09$).  Each model set differs from others in the physics inputs 
they employ, and these differences can result in systematic variations in the 
ages that are derived. As a result, it is worth identifying the factors that 
are most likely to affect the comparison between models and the stars in the 
WOCS 40007 binary and those that will affect the overall CMD comparison. 

For example, convective core overshoot (CCO) is believed to occur for
stars with $\mm \gtrsim 1.1~\msun$ \citep[the exact value depends on
the input chemical composition;][]{gallart}, and so it is expected
to be acting in the primary star of WOCS 40007 and possibly also the
secondary star. However, the effects of CCO are not clearly
identifiable until a star is near the cluster turnoff when CCO can
mean the difference between continued core fusion and central hydrogen
exhaustion. That translates into a large structural (radius)
difference between models that have different degrees of CCO. Thus CCO
will primarily affect comparisons in the CMD near the cluster turnoff.

CCO is handled in similar ways for all isochrone sets except for VR.  VR
isochrones use a parameterized version of the Roxburgh criterion
\citep[see][]{roxburgh,rosvick} and integrates fluid dynamics equations using
observationally determined parameters to constrain the amount of CCO present.
BaSTI, DSEP, and Y$^2$ force the amount of CCO to increase as the stellar mass
increases, using
  \[ \lambda_{OV}=\Lambda_{OS}H_p \]
where $\lambda_{OV}$ is the distance convective cells travel beyond the edge
of the convective zone as defined by the Schwarzschild criterion,
$\Lambda_{OS}$ is a characteristic CCO parameter, and $H_p$ is the local
pressure scale height. The precise shape of the isochrone is affected by how
$\Lambda_{OS}$ varies as a function of stellar mass.  We refer the reader to
the original papers for a full description of the CCO algorithm adopted by
each isochrone set.  However, to get an idea how this might influence the models, 
we compare below the amount of CCO used in each
of the isochrone sets for stars with masses at the cluster turnoff of NGC 6819
\citep[][$\sim1.47\msun$]{sandquist13} or like those stars in the WOCS 40007
binary ($1.24$ and $1.09 \msun$).

In the BaSTI isochrones \citep{pietrinferni}, $\Lambda_{OS}$ is
dependent on mass but not on metallicity, and generally lower than for
the other isochrone sets: $0.14$ at the turnoff, and 0.09 and 0.05 for
the binary star masses.  For Y$^2$ isochrones \citep{demarque},
$\Lambda_{OS} = 0$ and 0.05, respectively, for the components of WOCS
40007 because the onset of CCO is triggered at a relatively high mass
in those models.  However, $\Lambda_{OS}$ has reached its maximum
value (0.20) for the turnoff of the cluster.  DSEP parameterizes CCO
by adopting the same $\Lambda_{OS}$ treatment as Y$^2$, and the values
for the WOCS 40007 components and the cluster turnoff appear to be the
same. It should be remembered, however, that some CCO may be the result of
rotationally induced mixing, which is not taken into account by any of these
models.

The rate of the $^{14}$N$(p,\gamma)^{15}$O nuclear reaction rate is
important because it sets both the strength of the CNO cycle and the
extent of the convective core, and it has seen significant revision in
recent years \citep{marta}. Current public isochrone sets generally use older
and larger values for the rate. The exception is DSEP, which uses an 
earlier rate coefficient \citep{imbriani} that is
close to the most recent value. This would tend to reduce the extent
of core convection in the DSEP isochrones relative to other
sets. Helium diffusion (which is only employed in DSEP and Y$^2$
isochrones) can also potentially have an effect on the rate at which
helium builds up in the cores of stars.

Although the two eclipsing stars in WOCS 40007 have evolved significantly
in radius compared to zero-age stars, they are still on the
main sequence somewhat below the cluster turnoff and their
characteristics will not be as sensitive to age as stars nearer the
turnoff \citep[such as WOCS 23009;][]{sandquist13}. 

\subsection{Age of NGC 6819}

\subsubsection{CMD Comparison}\label{seccmdiso}
\begin{figure}
\begin{center}
\includegraphics[width = .9\textwidth]{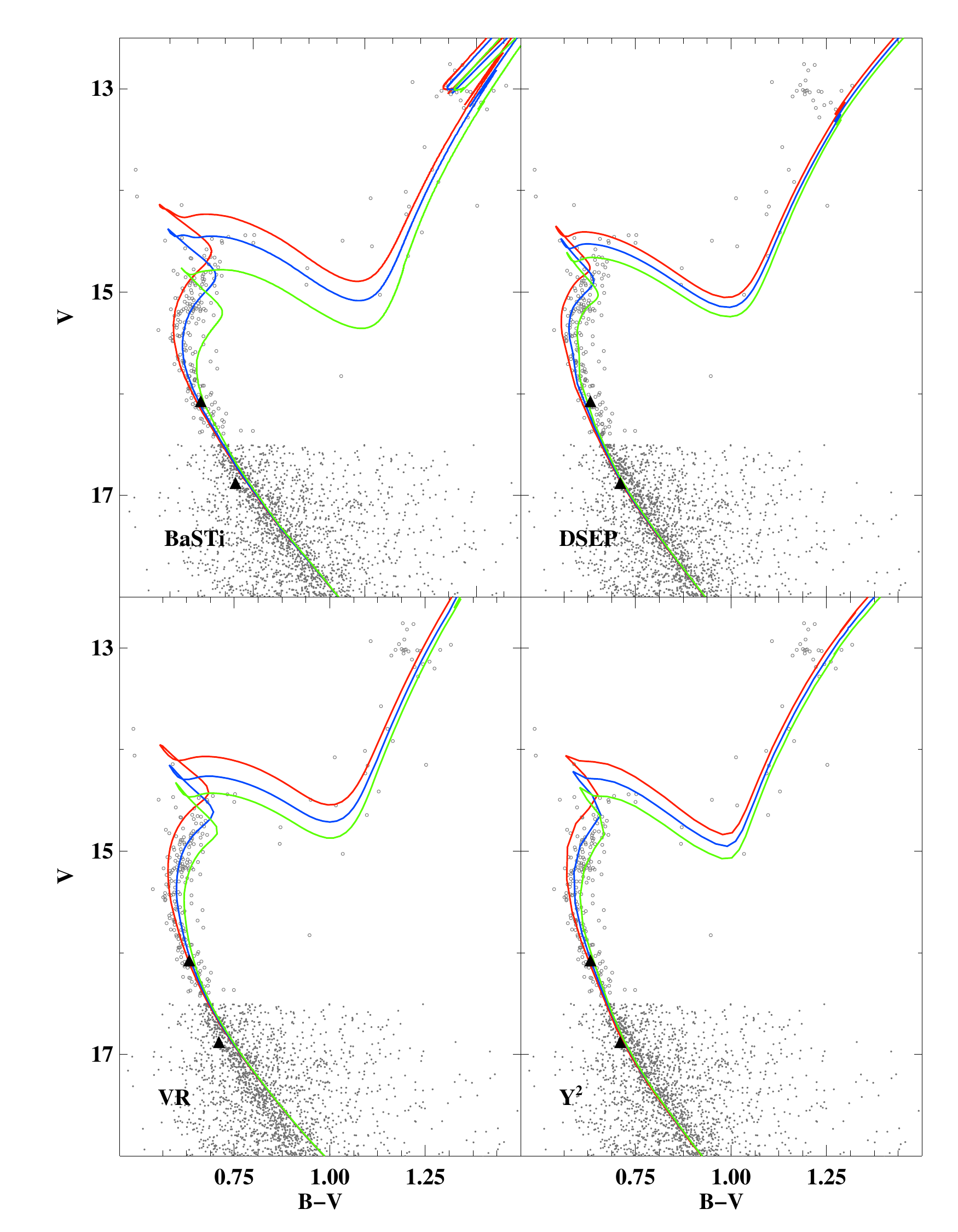}
\caption{ Color-magnitude diagram for NGC 6819 with 2.00, 2.25 and 2.50
  Gyr isochrones for BaSTI, VR and Y$^2$; and 2.50, 2.75, and 3.00 Gyr
  isochrones for DSEP. Isochrones were shifted using our adopted reddening of
  $E(B-V)=0.12$ and distance modulus $(m-M)_V=12.30$. The decomposed
  photometry of the WOCS 40007 stars is shown as solid
  points. \citet{kalirai} photometry is employed for the $(B-V,V)$ CMDs, while
  \citet{talamantes} photometry is used for $(V-I,V)$.
\label{cmdiso}}
\end{center}
\end{figure}

As a first check, we investigate how each isochrone set matches the CMD of NGC
6819 in the traditional way, using the reddening and distance determined in
Section \ref{reddist}.  In Figure \ref{cmdiso}, we plot the CMD with probable
single-star cluster members brighter than the $V\sim16.5$ magnitude
limit of the WOCS RV survey \citep{hole}.  Below this magnitude limit,
we plot all measured stars in $B-V$ (data from \citealt{kalirai}) and $V-I_c$
\citep{talamantes}.  The presence of CCO in stars near the turnoff of NGC 6819
is suggested by the cluster members near $(B-V,V)=(0.6,~14.6)$.

The BaSTI isochrones were generated using a metallicity [Fe/H] $= +0.06$,
which was the tabulated value closest to the spectroscopic measurement for NGC
6819.  The theoretical MS is slightly redder than the
cluster main sequence in the $(B-V)$ CMD, but the isochrones agree
reasonably well with the CCO-sensitive stars of the blue hook at the turnoff.
DSEP isochrones, made with [Fe/H] $= +0.09$,  produce the best overall
fit to the CMD, and the MS of the isochrones and CMD are well aligned.

VR isochrones were created with [Fe/H]$=+0.13$, which again was the tabulated
value closest to the spectroscopic measurement.  This isochrone set produces a
giant branch that is slightly too blue and misses a majority of the MS by
being too red, likely in part because the isochrone metallicity is larger than
the spectroscopic value for the cluster. In the vicinity of the turnoff, the
isochrones are generally too bright and/or red.
In the Y$^2$ models, the giant branch is again slightly too blue in $(B-V)$, but the MS fits well.
photometry in $(B-V)$. Strangely, the isochrones show much worse agreement 
on the main sequence and at the turnoff in $(V-I)$.

\begin{figure}
\begin{center}
\includegraphics[width = .9\textwidth]{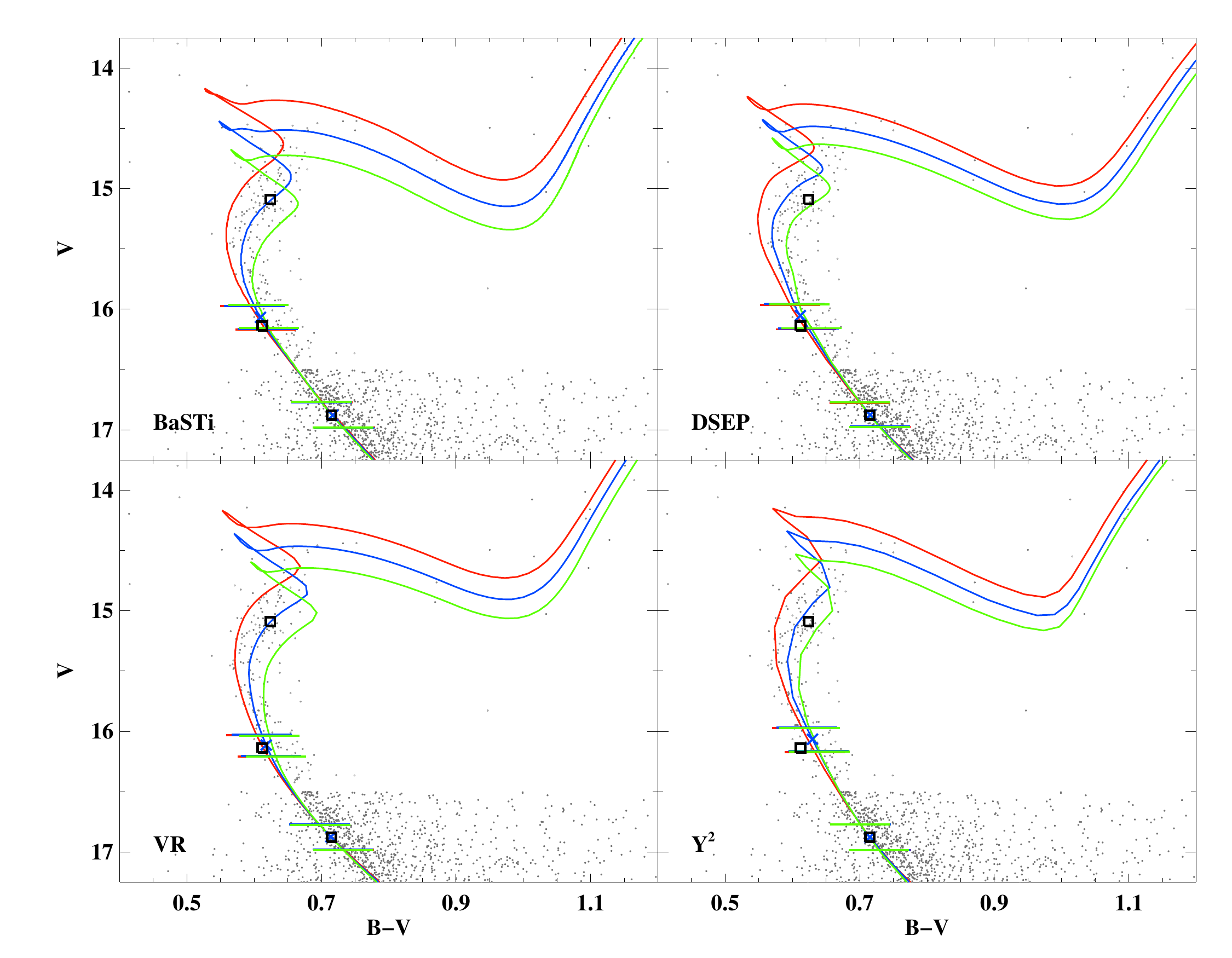}
\caption{ Color-magnitude diagram for NGC 6819 with isochrones shifted to fit
  the secondary star, as discussed in \S \ref{seccmdiso}. We also include
  the primary component of WOCS 23009, as determined by \citet{sandquist13}.
  Ages are 2.00, 2.25
  and 2.50 Gyr for BaSTI isochrones; and 2.25, 2.50 and 2.75 Gyr for DSEP, VR,
  and Y$^2$ isochrones.  Open squares show the decomposed photometry of WOCS
  40007 stars from \citet{kalirai} photometry, $\times$ symbols mark the
  isochrone value for the measured mass of each component, and horizontal
  lines delimit the range covered $1\sigma$ mass uncertainties. Green lines
  correspond to the greatest age, blue to the middle, and red to the lowest
  age.
\label{cmdage}}
\end{center}
\end{figure}

Given that there is uncertainty in the distance and reddening for the cluster,
we examine a second way of using isochrones by aligning them in the CMD with
the binary components at the observed mass values.  We use the $B-V$
decomposition of WOCS 40007 from the \citet{kalirai} photometry because it had
the highest signal-to-noise ratio for MS stars in the photometric
datasets. Because the photometry of the secondary component is independent of
assumptions about the tertiary star, we choose to force the isochrones to match
the secondary star's mass, $V$ magnitude, and $B-V$ color.  With this
alignment, the primary component lies within the range predicted by the
isochrones, although in some cases it lies near the lower photometric limit
(see Figure \ref{cmdage}).

With this type of fit, age information comes exclusively from the photometry
of cluster stars at the main sequence turnoff.  Given that the components of
WOCS 40007 fall near the blue edge of the main sequence, a good isochrone fit
should follow that edge. BaSTI isochrones do a decent job predicting the
location of the giant branch, but fail to properly fit the turnoff and
subgiant branch simultaneously. The 2.0 Gyr isochrone displays the best fit
to the turnoff, but has a subgiant branch that is too bright. As mentioned
earlier, this disagreement is probably the result of the smaller amount of CCO
used in these models.  DSEP isochrones fit well throughout the main sequence,
turnoff, subgiant branch, and red giant branch for an age of 2.5 Gyr.  For the
VR isochrones, the turnoff is fit for an age of about 2.3 Gyr, but the
subgiant branch is again too bright. Lastly, the Y$^2$ isochrones align well
with the giant branch and the turnoff, leading to a best fit age of slightly
below 2.5 Gyr.  Table \ref{tableages} lists the best fit age for each isochrone
in the ``CMD Isochrone'' column.

When fitting the isochrone to the binary components, there is an implied
reddening and distance modulus.  
The distance derived from the fit to the secondary star agrees with our
determination from the binary star components in Section \ref{secdm}
($(m-M)_V=12.44\pm0.07$) to within about $1\sigma$ for the BaSTI
[$(m-M)_V=12.35$], DSEP (12.40), and VR (12.36) isochrone sets, while the
Y$^2$ value (12.28) differs by slightly more than $2\sigma$ . There is
additional uncertainty in values derived by fitting to the primary star due to
poor constraints on the properties of the tertiary star, and generally the
distance moduli are larger (12.45, 12.51, 12.41, and 12.39 for BaSTI, DSEP,
VR, and Y$^2$, respectively). The difference is most likely a result of
uncertainties in the photometric decomposition, but additional effort to
detect and characterize the tertiary star would settle the question.

\subsubsection{Ages from the Mass-Radius Plane}

We compare the measured masses and radii of the known bright binary
components with model predictions. Ultimately we wish to use all of measurable
cluster stars for an analysis of the cluster age using this technique. For the
analysis below, we are including the primary star of the WOCS 23009 system
\citep{sandquist13}, although it should be remembered that WOCS 23009 is a
single-lined spectroscopic binary. As such, the primary star's mass had to be
estimated using the masses and $V$ magnitudes of WOCS 40007 in concert with
theoretical models. Because the estimated mass for that star varied from model
to model due to differences in the adopted physics, the position of the
error box varies slightly in these diagrams. 

\begin{figure}
\begin{center}
\includegraphics[width = \textwidth]{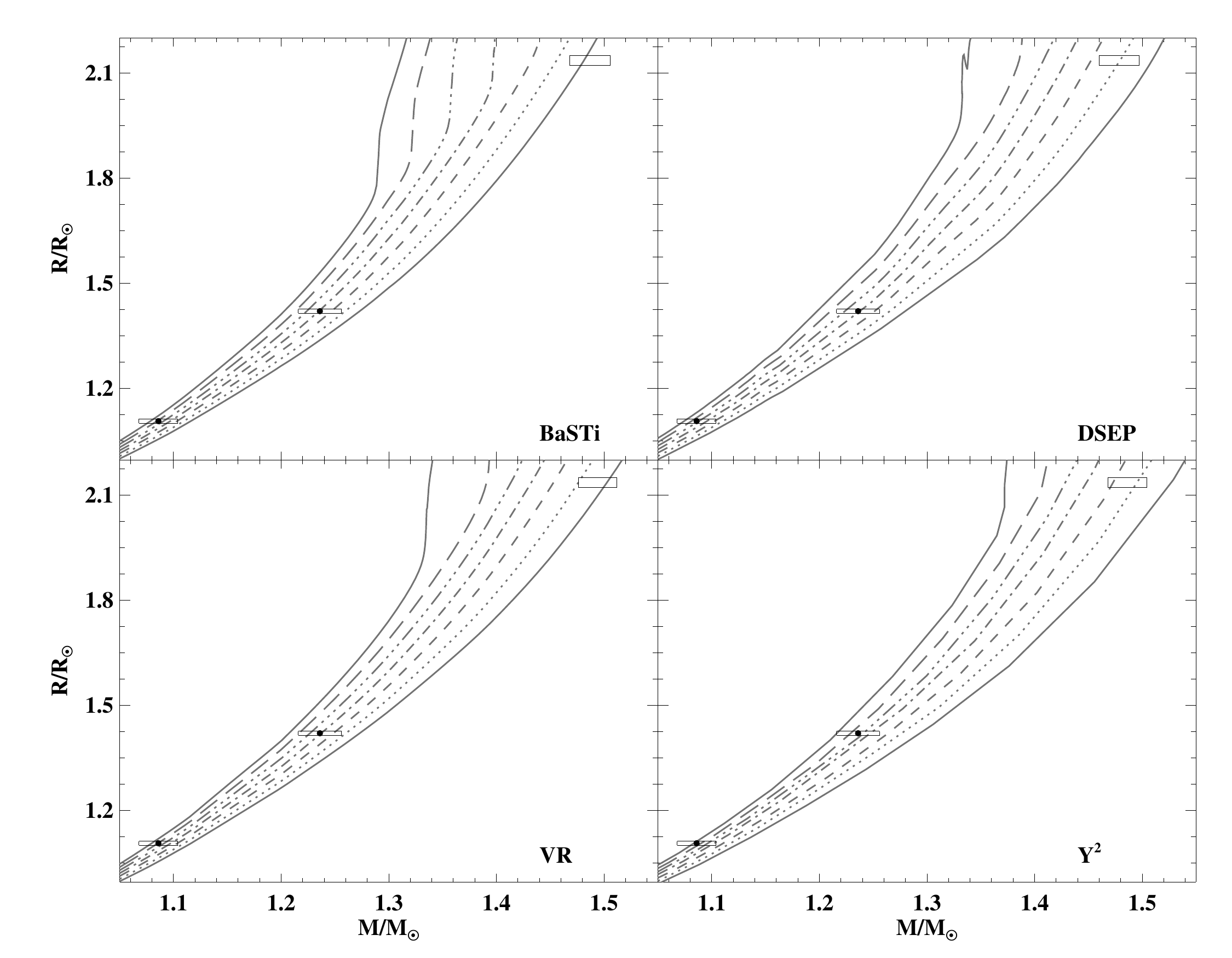}
\caption{
Derived masses and radii for the components of WOCS 40007 compared
with the four isochrone sets.  The isochrones employ
metallicites of $[{\rm Fe}/{\rm H}] = +0.06,~+0.09,~+0.13$ and 
+0.09 for BaSTI, DSEP, VR and Y$^2$ models, respectively.  Each panel has 
seven ages plotted and, from right to left, are: 2.25, 2.50, 2.75, 
3.00, 3.25, 3.50, and 3.75 Gyr. The error boxes correspond to a 
$1\sigma$ uncertainty.
}
\label{figiso}
\end{center}
\end{figure}

$M-R$ diagrams for each isochrone set are shown in Figure \ref{figiso}.  The
error boxes for the components of WOCS 40007 correspond to 1$\sigma$
uncertainties determined in \S \ref{elcmod}.  The radius uncertainties are
relatively small because the secondary eclipse is total.  Thus, it is the
precision of the mass measurements that currently determines the attainable
age precision, and the mass precision could be improved with additional, more
precise radial velocity measurements and an improved determination of the
tertiary star orbit.  Table \ref{tableages} shows the ages that we derive from
the respective $M-R$ isochrone fits from Figure \ref{figiso}.  The ages
derived from BaSTI, DSEP, and VR models using only the components of WOCS
40007 agree very well, while the Y$^2$ age is systematically higher. 

The characteristics of the primary star indicate an age of $2.9-3.3$ Gyr from
different isochrones with fitting uncertainties of 0.5 Gyr, while the
secondary returns $3.3-3.7$ Gyr with fitting uncertainties of 0.8 Gyr.  The
age determined from the secondary star is systematically higher, but there is
a range of isochrones that give a consistent age for both components within
the present $1\sigma$ mass uncertainties. Both stars of WOCS 40007 imply
greater ages than the more massive primary star of WOCS 23009, but the two
systems have consistent ages within $2\sigma$.  
If we take the measurements at face value though, the $M-R$
diagram analysis of WOCS 40007 produces a weighted average age estimate of
$3.1\pm0.4$ Gyr, where the statistical uncertainty is given. If the primary
star of WOCS 23009 is included in the weighted average, the age estimate
shifted significantly downward to $2.5\pm0.2$ Gyr.  In both cases, there is a
systematic uncertainty of at least 0.2 Gyr resulting from model uncertainties.

Based on evidence from field binary stars, there are reasons to
believe that the stars of WOCS 40007 could have somewhat anomalous radii.
Secondary stars of similar mass ($0.80-1.10\msun$) in field eclipsing binaries
with smaller periods ($0.6-2.8$ d) typically possess radii that are larger
than predictions and temperatures that are cooler \citep{clausen2009}.  The
cause is thought to be rotation-driven magnetic activity that inhibits
convective energy transport in the outer layers. WOCS 40007 has a period just
outside this interesting range, and magnetic activity would probably affect
the secondary to a greater degree due to its more massive convective
envelope. The primary component of WOCS 23009 is in a {\it much} longer period
binary and is likely to have an even smaller convective envelope, so that it
is unlikely to be affected by stellar activity. 
Regardless of the presence of magnetic activity, rotation in MS and turnoff
stars may cause an apparent shift of the stars in the CMD
\citep[][]{girardi2011}, but if the star in WOCS 23009 have synchronized
rotation, the rotation is probably too small to be a significant effect.  A
recent study by \citet{kaluzny} on the globular cluster M4 found that
different stars yield somewhat different ages and that the secondary stars
yielded larger age estimates, although at a statistically significant
level.
To more strongly test whether
all three stars are consistent with a single age or whether one or both of the
stars in WOCS 40007 are significantly inflated, we will need more precise
radial-velocity measurements in order to reduce the mass uncertainties.

\begin{figure}
\begin{center}
\includegraphics[width = \textwidth]{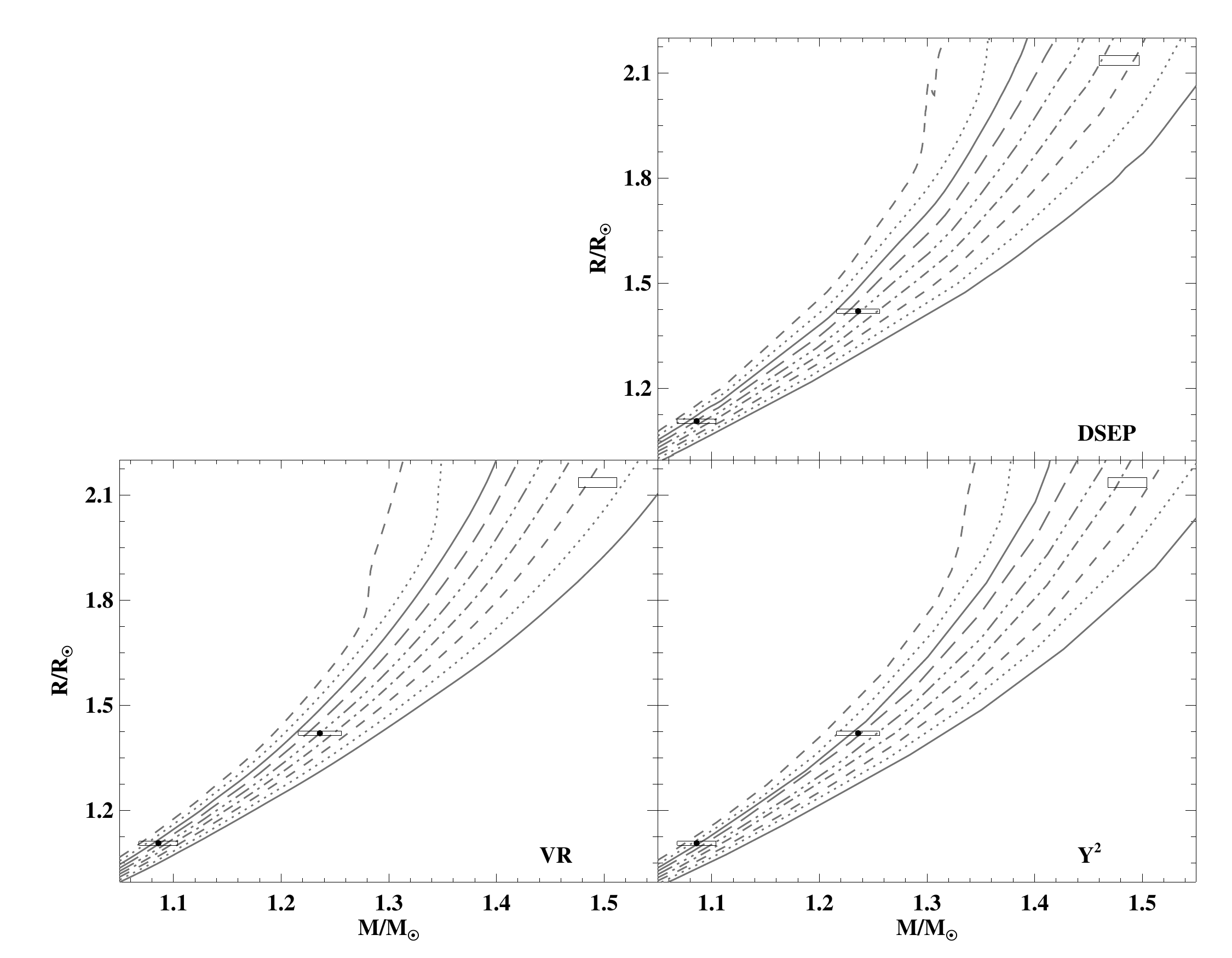}
\caption{
Derived masses and radii for the components of WOCS 40007 compared with three
of the four isochrone sets.  The isochrones were created with a decreased
metallicity value of $[{\rm Fe}/{\rm H}]$ = $-0.01$, 0.00 and $-0.01$ for
DSEP, VR and Y$^2$, respectively. (BaSTI isochrones are not included because
the tabulation does not include a closely matching metallicity value.)  Each
panel has nine ages plotted and, from right to left, are: 1.75, 2.00, 2.25,
2.50, 2.75, 3.00, 3.25, 3.50, and 3.75 Gyr. The error boxes correspond to a
1-$\sigma$ uncertainty.  }
\label{mrisocheck}
\end{center}
\end{figure}

\begin{figure}
\begin{center}
\includegraphics[width = .9\textwidth]{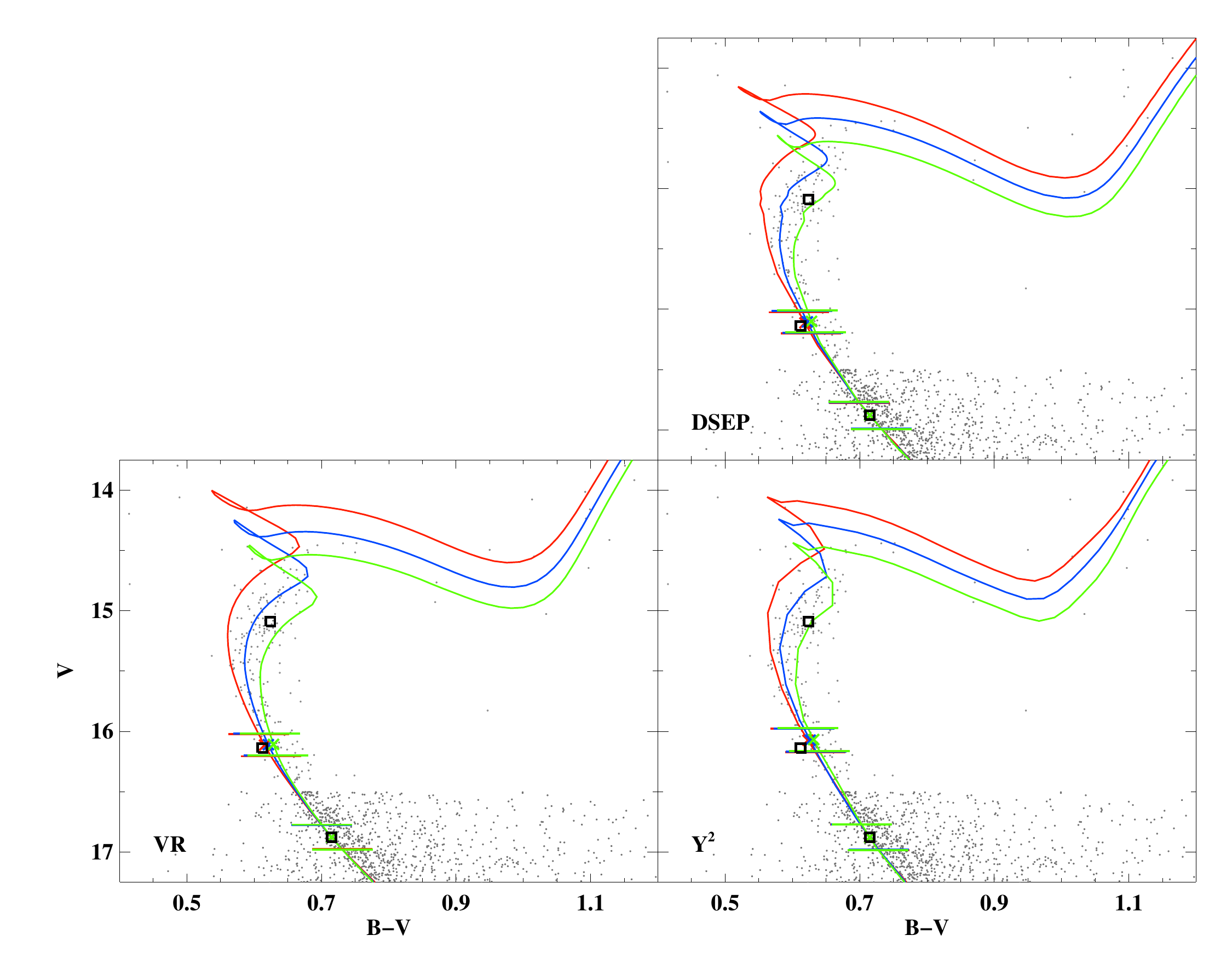}
\caption{
Color-magnitude diagram for NGC 6819 with 2.00, 2.25 and 2.50 Gyr isochrones
for DSEP, VR and Y$^2$.  Each isochrone set is plotted with a decreased
metallicity value of $[{\rm Fe}/{\rm H}] = -0.01$, 0.00, and $-0.01$ for DSEP,
VR, and Y$^2$, respectively. (BaSTI isochrones are not included because the
tabulation does not include a closely matching metallicity value.)  All
plotted points have the same definitions as in Figure \ref{cmdage}.  }
\label{cmdisocheck}
\end{center}
\end{figure}

The age estimate found from the analysis of the CMD in this and previous
studies of the cluster \citep{kalirai,rosvick} agrees best with the age
estimates from the primary component of WOCS 23009. While this may be an
indication that the stars in WOCS 40007 are inflated in size, isochrones in
the $M-R$ diagram shift significantly with chemical
composition. Therefore, we should consider the possibility that the adopted
metallicity may be incorrect or that another composition variable (like
helium; \citealt{brogaard}) may differ from what is assumed in the models.  We
tested this idea by changing the assumed [Fe/H] by $0.1$ dex, which is
significantly larger than the statistical uncertainty in the cluster
metallicity reported by \citet{bragaglia}, but within the range of
possibilities for typical systematic scale errors in abundance studies.
Figure \ref{mrisocheck} displays isochrones in the $M-R$ plane with a
metallicity decreased by 0.1 dex (which would make it closer in composition to
M67),
and Figure \ref{cmdisocheck} shows the same lower metallicity isochrones in
the CMD.  
We find that a decreased metallicity produces slightly greater consistency
between the ages implied by the two stars in WOCS 40007.  Additionally, the
lower metallicity produces ages are lower (by about 0.2 Gyr) from the stellar
masses and radii, and those ages are more consistent with those derived from
CMD fits.
The comparison of lower metallicity isochrones with the CMD returns ages of
$\sim2.25-2.50$ Gyr for the DSEP, VR, and Y$^2$ sets. (BaSTI is not considered
here because the tabulated isochrones with metallicity nearest to the reduced
value is $-0.25$, which is well below the prescribed 0.1 dex.)  In each case,
the age is more consistent with previously reported ages of the cluster. The
cluster turnoff is fit reasonably well by the lower metallicity isochrones,
although this is also influenced by the convective overshooting prescription.
Future spectroscopic work should be able to leverage the precise gravities
from the binary stars to improve the abundance analysis for stars at the
cluster turnoff and further investigate this possibility \citep{brogaard}.

\section{Summary \& Conclusions}
We have presented an extensive photometric and RV study of the detached
eclipsing binary star WOCS 40007 in NGC 6819 as part of a larger project to
precisely determine the age of the cluster. We use our precision mass and
radius determination measurements (uncertainties of $\sim1.6$\% and 0.5\%,
respectively) to derive an age of NGC 6819 of $3.1\pm0.4$ Gyr using BaSTI,
DSEP, and VR isochrones, while the best fit age for Y$^2$ isochrones is
slightly larger. The primary star of the brighter (and longer period) binary
WOCS 23009 indicates an age closer to 2.5 Gyr, which is more consistent with
the results from fitting isochrones in the CMD. However, the age results are
consistent to within $1-2 \sigma$ uncertainties.

The $M-R$ comparison is unaffected by many of the systematic uncertainties
that affect other methods of age determination (e.g., CMD isochrone fitting),
but is affected by chemical composition uncertainties.  The uncertainty in the
age estimate from the measured eclipsing stars is currently dominated by the
uncertainties in the stellar masses and the cluster chemical
composition. There is an indication that a slightly smaller metallicity would
produce greater consistency between mass-radius and CMD results. The
possibility of a lower cluster metallicity can still be spectroscopically
tested using the binary stars with the assistance of constraints on the
surface gravities from the binary-star analysis. In addition, more precise
mass measurements will help test whether the radii of the WOCS 40007 stars are
systematically high due as a consequence of their relatively short orbital
period.

With further work on known DEBs and forthcoming DEB discoveries in NGC 6819
from \kep, the age constraints from the $M-R$ diagram will be significantly
tightened.  An increased population of stars with well-determined mass, radius
and photometric properties will strongly constrain the fits in both
the $M-R$ diagram and the CMD, producing a situation where the input physics
of the isochrones may be thoroughly tested.

\acknowledgements We would like to thank the Director of Mount Laguna
Observatory (P. Etzel) for generous allocations of observing time and
A. Talamantes for his time and effort on observing and reducing a majority of
the $V$-band images at the MLO 1-meter.  Many thanks to E. Bavarsad and
D. Baer for their assistance with the observations at MLO.

This paper includes data collected by the \kep mission. Funding for the \kep
mission is provided by the NASA Science Mission directorate. We are grateful
to the \kep team for the opportunity to work with such a precise and extensive
photometric dataset. We gratefully acknowledge funding from the National
Science Foundation under grant AST-0908536 to San Diego State University and
AST-0908082 to the University of Wisconsin-Madison, and from the National
Aeronautics and Space Administration (NASA) under grant G00008957.  KB
acknowledges funding from the Carlsberg Foundation and the Villum Foundation.

\begin{deluxetable}{lcccc}
\tablecolumns{5}
\tablewidth{0pc}
\tablecaption{Photometric Observations at Mount Laguna Observatory 1-meter}
\tablehead{
\colhead{Date} & \colhead{Filters} & \colhead{mJD\tablenotemark{a} Start} &
\colhead{$t_{exp}$ (s)} & \colhead{N}}
\startdata
2010 Jun. 14 & $B$ & 5362.725 & 600 & 31 \\
2010 Jun. 22 & $B$ & 5370.708 & 600 & 37 \\
2010 Jun. 23 & $B$ & 5371.687 & 600 & 38 \\
2010 Jun. 25 & $V$ & 5373.680 & 600 & 34 \\
2010 Jul. 08 & $I_c$ & 5386.680 & 300 & 95 \\
2010 Jul. 19 & $I_c$ & 5397.664 & 300 & 100 \\
2010 Jul. 21 & $B$ & 5399.662 & 600 & 45 \\
2010 Aug. 20 & $R_c$ & 5429.649 & 420 & 60 \\
2010 Aug. 28 & $V$ & 5437.649 & 600 & 36 \\
2010 Aug. 31 & $R_cI_c$ & 5440.626 & 420, 300 & 36, 22 \\
2010 Sep. 01 & $B$ & 5441.824 & 600 & 13 \\
2010 Sep. 08 & $VR_c$ & 5448.618 & 600, 420 & 19, 16 \\
2010 Sep. 14\tablenotemark{b}&$BVR_c$&5454.780 & 600, 600, 480 & 7, 1, 7\\
2010 Sep. 17\tablenotemark{b} & $VR_cI_c$ & 5457.733 & 600, 480, 300 &
                                                                10, 10, 10 \\
2010 Sep. 21\tablenotemark{b} & $BVR_c$ & 5461.770 & 600, 600, 480 &
                                                                   5, 5, 5 \\
2010 Sep. 22\tablenotemark{b}&$BVR_cI_c$& 5462.604 & 600, 600, 480, 300&
                                                              8, 8, 13, 17 \\
2010 Oct. 02\tablenotemark{b}&$BVR_cI_c$& 5472.596 & 600, 600, 480, 300&
                                                                4, 2, 5, 6 \\
2010 Oct. 11\tablenotemark{b}&$BVR_cI_c$&5481.584 & 600, 600, 480, 300 &
                                                               8, 9, 7, 12 \\
2010 Oct. 17\tablenotemark{b} & $VR_cI_c$ & 5487.583 & 600, 480, 300 &  
                                                                  6, 9, 10 \\
2010 Oct. 30\tablenotemark{b} & $VR_cI_c$ & 5500.598 & 600, 480, 300 &
                                                                  8, 8, 12 \\
2010 Nov. 03 & $R_cI_c$ & 5504.581 & 480, 300 & 19, 16 \\
2010 Nov. 12 & $I_c$ & 5513.560 & 300 & 41 \\
2011 May  22 & $BV$ & 5704.801 & 600, 600 & 6, 14 \\
2011 Jun. 02 & $BV$ & 5715.738 & 600, 600 & 11, 13 \\
2011 Jun. 06 & $VI_c$ & 5719.722 & 600, 300 & 14, 14 \\
2011 Jun. 11 & $V$ & 5724.725 & 600 & 36 \\
2011 Jul. 03 & $I_c$ & 5746.663 & 300 & 12 \\
2011 Jul. 14 & $R_c$ & 5757.660 & 300 & 62 \\
2011 Jul. 23 & $I_c$ & 5766.657 & 300 & 83 \\
2011 Jul. 25 & $BI_c$ & 5768.661 & 600, 300 & 36, 41 \\
2011 Aug. 14 & $I_c$ & 5788.643 & 300 & 64 \\
2011 Aug. 25 & $BV$ & 5799.634 & 600, 600 & 17, 18 \\
2011 Sep. 16 & $BV$ & 5821.637 & 600, 600 & 17, 15
\enddata
\tablenotetext{a}{mJD = HJD - 2 450 000}
\tablenotetext{b}{Both twilight and dome flat fields used in the {\it IRAF}
reduction.}
\label{tableobs}
\end{deluxetable}

\begin{deluxetable}{lccccc}
\tablecolumns{6}
\tablewidth{0pc}
\tablecaption{Standard-system DEB Magnitudes and Effective Temperatures}
\tablehead{
\colhead{Name} & \colhead{$B$} & \colhead{$V$} & 
   \colhead{$R_c$} & \colhead{$I_c$} & 
   \colhead{T$_{\rm eff}$ (K)\tablenotemark{a}}}
\startdata
WOCS 40007 & $16.356\pm0.016$ & $15.651\pm0.016$ & & $14.774\pm0.018$ & \\
WOCS 40007$_{\rm p}$ & $16.739\pm0.017$ & $16.072\pm0.017$ & & $15.231\pm0.019$ & \\
WOCS 40007$_{\rm s}$ & $17.650\pm0.019$ & $16.896\pm0.019$ & & $15.944\pm0.021$ & \\
WOCS 40007$_3$ & $21.2\pm0.4$ & $19.8\pm0.4$ & & $18.3\pm0.4$ & \\
$\Delta m_{\rm s}$\tablenotemark{b} & $0.393\pm0.046$ & $0.415\pm0.024$ 
   & $0.402\pm0.032$ & $0.452\pm0.032$ & \\
\hline
\multicolumn{6}{c}{\citet{rosvick}}\\
WOCS 40007\tablenotemark{c} & $16.287\pm0.015$ & $15.622\pm0.010$ & & 
   $14.770\pm0.019$ & \\
WOCS 40007$_{\rm p}$ & $16.698\pm0.093$ & $16.072\pm0.039$ & & $15.291\pm0.068$ & \\
WOCS 40007$_{\rm s}$ & $17.581\pm0.214$ & $16.867\pm0.080$ & & $15.940\pm0.124$ & \\
\hline
\multicolumn{6}{c}{\citet{kalirai}}\\
WOCS 40007\tablenotemark{c} & $16.299$ & $15.633$ & & & \\
WOCS 40007$_{\rm p}$ & $16.705$ & $16.074$ & & & $6240\pm80$ \\
WOCS 40007$_{\rm s}$ & $17.593$ & $16.878$ & & & $5950\pm70$ \\
\hline
\multicolumn{6}{c}{\citet{hole} --- Phot03}\\
WOCS 40007\tablenotemark{c} & $16.294$ & $15.631$ & & & \\
WOCS 40007$_{\rm p}$ & $16.705$ & $16.081$ & & & \\
WOCS 40007$_{\rm s}$ & $17.588$ & $16.876$ & & & \\
\hline
\multicolumn{6}{c}{\citet{hole} --- Phot98}\\
WOCS 40007\tablenotemark{c} & $16.311$ & $15.657$ & $15.247$\tablenotemark{d} & 
   $14.824$ & \\
WOCS 40007$_{\rm p}$ & $16.722$ & $16.108$ & $15.874$ & $15.349$ & \\
WOCS 40007$_{\rm s}$ & $17.605$ & $16.902$ & $16.521$ & $15.994$ & \\
\enddata
\tablenotetext{a}{Temperature scale of \citet{casagrande}. See Section \ref{decomp} for a complete discussion.}
\tablenotetext{b}{$\Delta m_{\rm s}$ is the secondary eclipse depth 
relative to phases immediately before or after eclipse.}
\tablenotetext{c}{Calibrated system photometry has been shifted to the
  photometric zeropoints of this study.}
\tablenotetext{d}{\citet{hole} $R_c$ photometry has not been shifted to 
  our photometric zeropoints.}
\label{tablecomponents}
\end{deluxetable}

\begin{deluxetable}{lccccc}
\tabletypesize{\scriptsize}
\tablecolumns{5}
\tablewidth{0pc}
\tablecaption{Comparison of Photometric Zeropoints Relative to \citet{talamantes}}
\tablehead{
\colhead{Source} & \colhead{$\Delta B$} & \colhead{$\Delta V$} &
\colhead{$\Delta I_c$} & \colhead{$\Delta (B-V)$} & \colhead{$\Delta (V-I_c)$}}
\startdata
\citet{rosvick} & $-0.026$ (0.042) & $-0.015$ (0.033) & $-0.066$ (0.032)& $-0.011$ (0.023) & 0.049 (0.020)\\
\citet{kalirai} & $-0.046$ (0.033) & $-0.042$ (0.029) & & $-0.001$ (0.018) & \\
\citet{hole} Phot03 & $-0.026$ (0.021) & $-0.019$ (0.017) & & $-0.004$ (0.012) \\
\citet{hole} Phot98 & $-0.004$ (0.025) & $-0.001$ (0.027) & $-0.013$ (0.027) & $+0.002$ (0.018) & $+0.019$ (0.022) \\
\enddata
\tablecomments{Quoted values are for the median difference, with the semi-interquartile range in parentheses.}
\label{zpts}
\end{deluxetable}

\begin{deluxetable}{lcc}
\tablecolumns{3}
\tablewidth{0pc}
\tablecaption{Photometric and Spectroscopic Orbital Solutions of WOCS 40007}
\tablehead{
\colhead{Parameter} & \colhead{Limb-Darkening Coefficients} & 
  \colhead{Model Atmospheres}}
\startdata
\sidehead{Constrained Parameters:}
$e$ & 0 (fixed) & 0 (fixed)\\
$q~\left(\mm_{\rm 2}/\mm_{\rm 1}\right)$ & $0.878\pm0.009$ & $0.878\pm0.009$ \\
$K_1$ (km s$^{-1}$) & $89.6\pm0.5$ & $89.6\pm0.5$ \\
$K_2$ (km s$^{-1}$) & $102.0\pm0.6$ & $102.0\pm0.6$ \\
\sidehead{Free Parameters:}
P (d)\tablenotemark{a} & $3.1851308\pm0.0000012$ & $3.1851310\pm0.0000018$ \\
t$_0$ (${\rm HJD} - 2~450~000$) & $4698.7340\pm0.0003$ & $4698.7340\pm0.0004$\\
$i$ ($^o$) & $89.17\pm0.12$ & $88.81\pm0.06$ \\
$\gamma$ (km s$^{-1}$)\tablenotemark{b} & $0.62\pm0.32$ & $0.62\pm0.32$ \\ 
$\rr_{\rm 1}/a$ & $0.1164\pm0.0006$ & $0.1183\pm0.0004$ \\
$\rr_{\rm 1}/\rr_{\rm 2}$ & $1.274\pm0.006$ & $1.293\pm0.006$ \\
$T_{\rm 2}/T_{\rm 1}$ & $0.935\pm0.007$ & $0.936\pm0.001$ \\
\hline
\sidehead{Derived Parameters:}
$\mm_{\rm 1}$ ($\msun$) & $1.236\pm0.020$ & $1.236\pm0.020$ \\
$\mm_{\rm 2}$ ($\msun$) & $1.086\pm0.018$ & $1.086\pm0.018$ \\
$\rr_{\rm 1}$ ($\rsun$) & $1.399\pm0.007$ & $1.423\pm0.005$ \\
$\rr_{\rm 2}$ ($\rsun$) & $1.098\pm0.004$ & $1.100\pm0.004$ \\
$\rr_{\rm 2}/a$ & $0.0913\pm0.0003$ & $0.0915\pm0.0003$ \\
$(\rr_{\rm 1}+\rr_{\rm 2})/a$ & $0.2078\pm0.0007$ & $0.2098\pm0.0008$ \\
$a$ ($\rsun$) & $12.022\pm0.001$ & $12.023\pm0.001$ \\
log $g_{\rm 1}$ (cgs) & $4.240\pm0.005$ & $4.226\pm0.004$ \\
log $g_{\rm 2}$ (cgs) & $4.381\pm0.004$ & $4.379\pm0.004$ \\
$v_{rot,1}\sin i (\kms)$\tablenotemark{c} & $22.23\pm0.11$ & $22.61\pm0.08$\\
$v_{rot,2}\sin i (\kms)$\tablenotemark{c} & $17.45\pm0.07$ & $17.48\pm0.07$\\
\enddata
\tablenotetext{a}{Period derived strictly from ground-based data, and is
not corrected for effects of the tertiary star.}
\tablenotetext{b}{We assume that the recessional velocity is identical
   for both components.}
\tablenotetext{c}{Computed assuming synchronism.}
\label{tabsolutions}
\end{deluxetable}

\begin{deluxetable}{lccl}
\tablewidth{0pc}
\tablecaption{Eclipse Timings for WOCS 40007}
\tablehead{
\colhead{BJD-2450000} & \colhead{$\sigma$} & \colhead{Eclipse} & \colhead{Notes}}
\startdata
2102.88906 & 0.00042 & P & ground $R_C$ \\
4623.88256 & 0.00024 & S & ground $V$ \\
4698.73159 & 0.00020 & P & ground $R_K$ \\
4964.69070 & 0.00025 & S & start of Kepler Q1 \\
4966.28308 & 0.00023 & P & \\
4967.87586 & 0.00019 & S & \\
4969.46819 & 0.00013 & P & \\
4971.06104 & 0.00005 & S & \\ 
4972.65339 & 0.00012 & P & \\
4974.24608 & 0.00022 & S & \\ 
4975.83853 & 0.00022 & P & \\
\enddata
\tablecomments{This table is published in its entirety in the electronic edition, 
but a portion is shown here for guidance regarding its form and content.}
\label{etiming}
\end{deluxetable}

\begin{deluxetable}{lcc}
\tablecolumns{3}
\tablewidth{0pc}
\tablecaption{Orbit Solution for the Tertiary Star in WOCS 40007}
\tablehead{\colhead{Parameter} & \colhead{Value}}
\startdata
$P_{\rm b}$ (d) & $3.18509876\pm0.00000007$ \\
$P_{\rm 3}$ (d) & $3547\pm33$ \\
$t_{\rm b}$ & $2452102.8797\pm0.0001$ \\
$t_{\rm 3}$ & $2455613\pm5$ \\
$e_{\rm 3}$ & $0.157\pm0.002$ \\
$\omega_{\rm 3}$ ($\degr$) & $77.0\pm0.7$ \\
$K_1$ (km s$^{-1}$) & $89.5\pm0.7$ \\
$K_2$ (km s$^{-1}$) & $101.9\pm0.8$ \\
$K_3$ (km s$^{-1}$) & $3.92\pm0.03$ \\
$\gamma$ (km s$^{-1}$) & $0.62\pm0.32$ \\
$q=\mm_{\rm 2}/\mm_{\rm 1}$ & $0.879\pm0.010$ \\
\enddata
\label{terttab}
\end{deluxetable}

\clearpage

\begin{deluxetable}{lcccc}
\tablecolumns{4}
\tablewidth{0pc}
\tablecaption{Age Estimates for NGC 6819}
\tablehead{
\colhead{Model} & \colhead{CMD} & \multicolumn{3}{c}{$M-R$ Diagram} \\
& & \multicolumn{2}{c}{WOCS 40007} & \colhead{WOCS 23009} \\
& \colhead{Isochrones} & \colhead{Primary} & \colhead{Secondary} &}
\startdata
BaSTI & $2.1$ & $2.9\pm0.5$ & $3.4\pm0.8$ & $2.2\pm0.2$\\
DSEP & $2.5$ & $3.0\pm0.5$ & $3.3\pm0.8$ & $2.5\pm0.2$ \\
VR  & $2.3$ & $3.0\pm0.5$ & $3.4\pm0.8$ & $2.3\pm0.2$ \\
Y$^2$ & $2.4$ &$3.3\pm0.5$ & $3.7\pm0.8$ & $2.6\pm0.2$\\ 
\enddata
\label{tableages}
\end{deluxetable}

\end{document}